\documentstyle[12pt]{article} 
\begin{document} \begin{titlepage}
\begin{tabbing}
   \hspace{13.4cm} \=    \kill \>   \=  IFP-470 \+\+ \\
	   July 1993
\end{tabbing} \begin{center}

\Large{{\bf Canonical variables and quasilocal}\\ {\bf energy in
general relativity}}

\bigskip \bigskip \normalsize Stephen Lau

\bigskip {\it Institute of Field Physics and\\ Theoretical Astrophysics
and Relativity Group\\ Department of Physics and Astronomy\\ The
University of North Carolina\\ Chapel Hill, NC 27599-3255\\}

\begin{abstract} Recently Brown and York have devised a new method for
defining quasilocal energy in general relativity. Their method employs
a Hamilton-Jacobi analysis of an action functional for a spatially
bounded spacetime $M$, and this analysis yields expressions for the
quasilocal energy and momentum surface densities associated with the
two-boundary $B$ of a spacelike slice of such a spacetime.  These
expressions are essentially Arnowitt-Deser-Misner variables, but with
canonical conjugacy defined with respect to the time history $^{3}\! B$
of the two-boundary.  These boundary ADM variables match previous
variables which have arisen directly in the study of black hole
thermodynamics. Further, a ``microcanonical" action which features
fixed-energy boundary conditions has been introduced in related work
concerning the functional integral representation of the density of
quantum states for such a bounded gravitational system. This paper
introduces Ashtekar-type variables on the time history $^{3}\! B$ of
the two-boundary and shows that these variables lead to elegant
alternative expressions for the quasilocal surface densities.  In
addition, it is demonstrated here that both the boundary ADM variables
and the boundary Ashtekar-type variables can be incorporated into a
larger framework by appealing to the tetrad-dependent Sparling
differential forms. Finally, using these results and a tetrad action
principle employed by Goldberg, this paper constructs two new tetrad
versions of the microcanonical action.  \end{abstract} \end{center}
\end{titlepage}

\section{Introduction}

Recent investigations in black hole physics have revealed the
fundamental role played by boundary conditions in the equilibrium
theory of self-gravitating systems. The following principal ideas have
spawned this new perspective: (1) equilibrium systems are necessarily
spatially finite when general relativity is taken into account, and (2)
the selection of a statistical ensemble dictates what boundary terms
are present in the classical action.\cite{BMYW} The various ensembles
corresponding to the same self-gravitating system differ by what
thermodynamic data are fixed on the spatial boundary $B$. When treating
a system in a particular ensemble and constructing the corresponding
partition function as a functional integral, one must use the classical
action associated with the appropriate variational principle.
Therefore, the importance of the spatial boundary and its time history
$^{3}\!B$ has been elevated. This history is a (2+1)-dimensional
Lorentz manifold. As a result of this new emphasis, attention has been
focused on the canonical structure associated with $^{3}\! B$.

Motivated by these considerations, Brown and York have devised a
somewhat different method for defining quasilocal energy in general
relativity.\cite{BY} The new method consists of a Hamilton-Jacobi
analysis of a classical action functional appropriate for such a
spatially bounded system. Though a sketch of this analysis is presented
below, the reader is urged to consult the original reference. Following
this method, one constructs expressions for the quasilocal energy,
quasilocal momentum, and quasilocal spatial stress surface densities
which are point-wise functions on the two-boundary $B$. These
expressions are intimately related to variables which arise directly in
the examination of self-gravitating thermodynamical systems. For
instance, the quasilocal energy, the integral over $B$ of the
quasilocal energy surface density, plays the role of the total
thermodynamical internal energy and is conjugate to inverse
temperature. It has been emphasized that thermodynamical conjugacy and
canonical conjugacy with respect to $^{3}\! B$ are essentially the
same.\cite{micro} Both types of conjugacy are present simultaneously,
and both illuminate different aspects of the gravitational field. In
this spirit the quasilocal expressions, which are basically the ADM
variables but defined on $^{3}\! B$ rather than a spacelike slice, may
be viewed as ``thermodynamical" variables.

This work demonstrates that new Ashtekar-type variables can be
introduced on $^{3}\!B$.  These {\em boundary Ashtekar variables} enjoy
the usual magical properties associated with the conventional Ashtekar
variables, and, furthermore, they allow an elegant reformulation of
some of the major Brown-York results. In particular they allow a
compact expression for the boundary piece of the Hamiltonian  which,
modulo the constraints of general relativity, is the full Hamiltonian
for such a spatially bounded system. Since one of the new boundary
variables is a connection, their utility seems quite promising. Indeed,
the existence of the boundary Ashtekar variables has already suggested
some interesting avenues of research in gravitational thermodynamics
which will be reported elsewhere. Also, this discussion shows that both
the original Brown-York variables and boundary Ashtekar variables can
be incorporated into a larger framework. This is achieved by appealing
to Cartan's calculus of orthonormal frames and to certain
frame-dependent differential forms, first discovered by Sparling, that
enjoy properties linking them closely with notions of gravitational
energy. Finally, this paper uses the developed formalism and a tetrad
action principle employed by Goldberg to construct two new tetrad
versions of the microcanonical action. The microcanonical action
features fixed-energy boundary conditions.  Brown and York have
presented the microcanonical action in a separate but related work,
where they have used it in a functional integral representation of the
density of quantum states associated with a spatially bounded
self-gravitating system. \cite{micro}

\section{Preliminaries}

This section collects some notational conventions and presents a
skeletal description of the Brown-York method for constructing the
quasilocal expressions. To facilitate the introduction of the boundary
Ashtekar variables, the Brown-York results are then presented in the
language of orthonormal frames. The conventions presented here adhere
almost fully to those found in \cite{BY}.

\subsection{Conventions and assumptions} Throughout this discussion the
focus of interest is a spacetime region $M$ which is topologically the
Cartesian product of a three-manifold $\Sigma$ and a closed connected
segment of the real line $I$. The spacetime $M$ is foliated by a time
function $t: M\rightarrow I$. The level hypersurfaces of this function
or ``slices" are the leaves of the foliation, and the slices labeled by
the initial and final endpoints of $I$ are denoted respectively by $t'$
and $t''$. The letter $\Sigma$ is used both to denote the foliation of
$M$ and to refer to a generic slice of this foliation.  The timelike,
future pointing, unit, hypersurface normal of the foliation $\Sigma$ is
denoted by $u$. Also it is assumed that the three-space $\Sigma$ has a
boundary $\partial\Sigma = B$ which is not necessarily simply
connected. The product of $B$ with $I$ is an element of the
three-boundary of $M$ and is denoted by $^{3}\!B$. This element is
often referred to as the three-boundary, even though technically the
three-boundary of $M$ consists of $^{3}\!B$, $t'$, and $t''$.  The
spacelike, outward pointing, unit, normal vector field of $^{3}\!B$ is
represented by $n$. Since $^{3}\!B$ is a $(2+1)$-dimensional spacetime
in its own right, one may consider the foliation of $^{3}\!B$ by
$t|_{^{3}\!B}$, the restriction of the time function $t$ to the
three-boundary, independently of the complete spacetime $M$ and its
foliation $\Sigma$. The letter $B$ is used both to denote the
independent foliation of $^{3}\!B$ and to refer to a generic leaf of
this foliation. Regarding the compatibility of the foliations $B$ and
$\Sigma$, the following crucial assumption is made: the leaves of the
hypersurface foliation $\Sigma$ of $M$ are orthogonal to $^{3}\!B$.
This condition requires that the hypersurface normal $u$ is orthogonal
to the three-boundary normal $n$. Equivalently, this condition implies
that the restriction of $u$ to $B = \partial\Sigma$ is the timelike,
future pointing, unit, normal vector field of the two-surface $B$ when
it is viewed as a particular leaf of the foliated $^{3}\!B$.

The induced metric tensor and extrinsic curvature tensor of a generic
slice $\Sigma$ as embedded in $M$ are denoted respectively by $h_{ij}$
and $K_{ij}$. Lower case Latin letters from the latter half of the
alphabet $(i,j,k,\cdots )$ are used as coordinate indices on $\Sigma$
and run over the values $(1, 2, 3)$. Letters of the same type with hats
are used as orthonormal indices and labels on a slice and run over the
values $(\hat{1},\hat{2},\hat{3})$.  The induced metric and extrinsic
curvature of $^{3}\!B$ as a manifold embedded in $M$ are denoted
respectively by $\gamma_{ij}$ and $\Theta_{ij}$. Lower case Latin
letters from the latter half of the alphabet are also used as
coordinate indices on $^{3}\!B$, however, $^{3}\!B$ coordinate indices
run over the values $(0, 1, 2)$. Again the same letters with hats refer
to an orthonormal frame, but on $^{3}\!B$ indices with hats take the
values $(\bot,\hat{1},\hat{2})$. For $B$, a generic leaf of the
foliated $^{3}\!B$, $\sigma_{ab}$ and $k_{ab}$ are used respectively as
the intrinsic metric tensor and extrinsic curvature tensor of the
two-surface as embedded in $\Sigma$.  One should note that
$\sigma_{ab}$ is also the induced metric of $B$ as a surface embedded
in $^{3}\! B$. There is no need here to consider the extrinsic
curvature tensor of $B$ as embedded in $^{3}\! B$. Coordinate indices
on $B$ are represented by lower case Latin letters from the beginning
of the alphabet $(a,b,c,\cdots ,h)$ and run over the values $(1,2)$,
while orthonormal indices on $B$ are represented by the same letters
with hats and run over $(\hat{1},\hat{2})$.  When potential confusion
arises between slice and three-boundary indices or between slice and
$B$ indices, the practice of underlining slice indices, {\em e.g.}
$h_{\underline{ij}}$, is followed. The metric on $M$ is denoted by
$g_{\mu\nu}$. Spacetime coordinate indices are represented by lower
case Greek letters $(\mu,\nu,\sigma,\cdots )$ and run over the values
$(0,1,2,3)$, while orthonormal indices on $M$ are represented by the
same letters with hats and run over $(\bot,\hat{1},\hat{2},\hat{3})$.
Finally, $\epsilon_{\hat{r}\hat{s}\hat{t}}$ is used to denote the
orthonormal components of the Levi-Civita tensor on both $\Sigma$ and
$^{3}\!B$. On $\Sigma$ these components are determined by
$\epsilon_{\hat{1}\hat{2}\hat{3}} = 1
=\epsilon^{\hat{1}\hat{2}\hat{3}}$, while on $^{3}\!B$ one has
$\epsilon_{\bot\hat{1}\hat{2}} = 1 = -\epsilon^{\bot\hat{1}\hat{2}}$.
The orthonormal Levi-Civita tensor on $B$ is denoted by
$\epsilon_{\hat{a}\hat{b}}$ with $\epsilon_{\hat{1}\hat{2}} = 1 =
\epsilon^{\hat{1}\hat{2}}$, and on $M$ it is denoted by
$\epsilon_{\hat{\alpha}\hat{\beta}\hat{\mu}\hat{\nu}}$ with
$\epsilon_{\bot\hat{1}\hat{2}\hat{3}} = 1 = -
\epsilon^{\bot\hat{1}\hat{2}\hat{3}}$.

A variety of derivative operators is used in this work. The ordinary
metric compatible covariant derivative on a generic slice $\Sigma$
which acts on coordinate indices is denoted by $D_{i}$. For a vector
field $v = v^{i}\partial_{i}$ on $\Sigma$ its action is given by

\begin{equation} D_{j}\,v^{i} = \partial_{j}\,v^{i} +
v^{k}\,\Gamma^{i}_{kj}\, , \end{equation} where the $\Gamma^{i}_{kj}$
are the ordinary Christoffel symbols on $\Sigma$ determined by
$h_{ij}$. In addition to $D_{i}$ there exists the orthonormal frame
representation of this derivative which acts on orthonormal $\Sigma$
indices, and this operator is denoted by $\overline{D}_{i}$. An
orthonormal frame or {\em triad} $\left\{ E_{\hat{s}} =
E_{\hat{s}}\,^{j}\partial_{j} | \hat{s} =
\hat{1},\hat{2},\hat{3}\right\}$ determines connection coefficients
$\omega^{\hat{r}}\,_{\hat{s}\hat{t}} =
\omega^{\hat{r}}\,_{\hat{s}j}E_{\hat{t}}\,^{j}$.  For the same vector
field $v$ expressed as $v = v^{\hat{s}}\,E_{\hat{s}}$, the action of
$\overline{D}_{j}$ is defined by

\begin{equation} \overline{D}_{j}\,v^{\hat{s}} =
\partial_{j}\,v^{\hat{s}} +
v^{\hat{t}}\,\omega^{\hat{s}}\,_{\hat{t}j}\, .  \end{equation} The
analogous operators on the three-boundary $^{3}\!B$ are represented by
${\cal D}_{i}$ and $\overline{{\cal D}}_{i}$. The analogous operators
on $B$ are denoted by $^{2}\nabla_{b}$ and $^{2}\overline{\nabla}_{b}$
and on spacetime $M$ by $\nabla_{\mu}$ and $\overline{\nabla}_{\mu}$.
Finally, $\hat{{\cal D}}_{i}$ is used for the Ashtekar derivative on
$^{3}\!B$. The major conventions of this paper are collected in a table
found on the last page. All remaining conventions used in this analysis
are are presented ``locally" throughout the text.

\subsection{Review of the Brown-York method}

The analysis of Brown and York \cite{BY} begins with the action for
pure Einstein gravity which is appropriate for a variational principle
in which the induced metric is fixed on all the elements of the
three-boundary. Namely,\cite{York}

\begin{equation} S^{1} =
\frac{1}{2\kappa}\int_{M}d^{4}x\,\sqrt{-g}\,\Re +
\frac{1}{\kappa}\int_{t'}^{t''}d^{3}x\,\sqrt{h}\,K -
\frac{1}{\kappa}\int_{^{3}\!B}d^{3}x\,\sqrt{-\gamma}\,\Theta .
\label{BYaction1}\end{equation}
 In the above formula $\Re$ is the Ricci scalar associated with the
 metric space $M$, and the notation $\int_{t'}^{t''}d^{3}x$ is a
compact expression for $\int_{t''}d^{3}x - \int_{t'}d^{3}x$.  The
variation of this action is given by

\begin{eqnarray} \delta S^{1} & = & ( {\rm terms\, giving\,the\,
equations\, of\, motion} )   \\
	 &   &  + \int_{t'}^{t''}d^{3}x\,p^{ij}\,\delta h_{ij} +
\int_{B^{3}}d^{3}x\,\pi^{ij}\,\delta\gamma_{ij}\, .  \nonumber
\end{eqnarray} In the above expression $p^{ij}$ denotes the ADM
gravitational momentum conjugate to $h_{ij}$

\begin{equation} p^{ij} = \frac{1}{2\kappa}\sqrt{h}\left( Kh^{ij} -
K^{ij}\right)\, .  \label{slicemomentum1} \end{equation} Likewise,
$\pi^{ij}$ represents the gravitational momentum conjugate to
$\gamma_{ij}$, where in this case conjugacy is defined with respect to
$^{3}\!B$. One finds that

\begin{equation} \pi^{ij} = - \frac{1}{2\kappa}\sqrt{-\gamma}\left(
\Theta\gamma^{ij} - \Theta^{ij}\right) .  \label{bdmomentum1}
\end{equation} As described in \cite{BY}, it is often necessary to
consider actions which differ from (\ref{BYaction1}) by a subtraction
term $S^{0}$ which is a functional solely of the fixed boundary data
$\gamma_{ij}$. The complete action is therefore given by

\begin{equation} S = S^{1} - S^{0} .      \label{BYaction2}
\end{equation}

If the general variation of the action $S^{1}$ is restricted to
variation among just the classical solutions, then the action
(\ref{BYaction1}) is identified with the Hamilton-Jacobi principal
function $S^{1}_{c\ell}$. The subscript $c\ell$ denotes ``evaluation on
a classical solution." The principal function $S^{1}_{c\ell}$ is a
functional of the fixed boundary data $\gamma_{ij}, h'_{ij}, h''_{ij}$.
Clearly there are only boundary contributions to $\delta
S^{1}_{c\ell}$. Since the fixed three-boundary metric $\gamma_{ij}$
provides the information necessary to determine the lapse of proper
time between the hypersurface slices $t'$ and $t''$, in analogy with
Hamilton-Jacobi theory for simple systems one would expect that
$\pi^{ij}_{c\ell}$, the gravitational momentum determined by a
classical history, is related to a gravitational energy density.
However, since $\gamma_{ij}$ also contains all the metrical information
about $^{3}\!B$, the momentum $\pi^{ij}_ {c\ell}$ provides more than
just a quasilocal energy density. The three-boundary momentum may also
be used to define a quasilocal momentum density and quasilocal spatial
stress density. The argument of this paragraph applies equally well to
the action (\ref{BYaction2}), however, in this case one must consider
an additional term to $\pi_{c\ell}^{ij}$ given by $\delta
S^{0}/\delta\gamma_{ij}|_{c\ell}$ . Therefore, the quasilocal
expressions are not unique.

To construct the quasilocal expressions a (2+1)-decomposition of the
metric on $^{3}\!B$ must first be performed.  In coordinates adapted to
the foliation the metric on $^{3}\!B$ takes the ADM form

\begin{equation} \gamma_{ij}\, dx^{i}\otimes dx^{j} = -N dt\otimes N dt
+ \sigma_{ab}\left( dx^{a} + V^{a}dt\right)\otimes \left( dx^{b} +
V^{b}dt\right)\, . \label{bdmetric} \end{equation} The lapse function
$N$ and shift vector field $\vec{V}$ are usually kinematical variables,
but here they constitute part of the fixed boundary data on $^{3}\!B$.
Following Brown and York, one defines the quasilocal energy and
momentum surface densities by

\begin{eqnarray}
  &  &  \varepsilon\, :=\, -\frac{1}{\sqrt{\sigma}}\frac{\delta
  S_{c\ell}}{\delta N} = \left.\frac{1}{\kappa}\, k \right|_{c\ell}  +
\frac{1}{\sqrt{\sigma}}\frac{\delta S^{0}_{c\ell}}{\delta N}
\label{ep}\\
  &  &  j_{b}\, :=\, \frac{1}{\sqrt{\sigma}}\frac{\delta
  S_{c\ell}}{\delta V^{b}} = \left.- \frac{2}{\sqrt{h}}\,
\sigma_{b\underline{i}}\, n_{\underline{j}}\, p^{\underline{ij}}
\right|_{c\ell} - \frac{1}{\sqrt{\sigma}}\frac{\delta
S^{0}_{c\ell}}{\delta V^{b}}\, , \label{jay} \end{eqnarray} where
$\sqrt{\sigma}$ denotes the square root of the determinant of the
two-metric of $B$.  The final expressions for the quasilocal energy and
momentum densities are discussed further below.  For the above
definitions it is quintessential to note that these functional
derivatives are with respect to the lapse and shift {\em on the
three-boundary} $^{3}\!B$ which, as mentioned, are not considered
kinematical variables. Since it is not needed in this work, the form of
the spatial stress density is not presented above. If the subtraction
term $S^{0}$ is functionally linear in $N$ and $\vec{V}$, then the
above expressions depend only on slice Cauchy data. In this case
$\varepsilon$ and $j_{b}$ are functionals on the $\Sigma$ phase space.
Therefore, the $c\ell$ in the above expressions may be omitted with the
understanding that these expressions are to be evaluated on a point of
phase space $\left(h_{ij}, p^{ij}\right)$ corresponding to a solution
of the equations of motion.\cite{BY} This viewpoint is adopted in what
follows.

\subsection{Triad version of the Brown-York method}

In order to pass to a connection formalism it is necessary to enlarge
the phase space associated with the boundary.\cite{Ashtekar1} This is
achieved by the introduction of an orthonormal frame (technically
pseudo-orthonormal) or triad $\left\{ \xi_{\hat{r}} \mid \hat{r} =
\bot, \hat{1}, \hat{2}\right\}$ on $^{3}\!B$ and its associated cotriad
$\left\{ \xi^{\hat{r}}  \mid \hat{r} = \bot,\hat{1},\hat{2}\right\}$.
In a coordinate system the frame and coframe may be expanded
respectively as $\xi_{\hat{r}} = \xi_{\hat{r}}\,^{j}\partial_{j}$ and
$\xi^{\hat{r}} = \xi^{\hat{r}}\,_{j} dx^{j}$. Orthonormality implies
that $\eta_{\hat{r}\hat{s}} =
\gamma_{ij}\,\xi_{\hat{r}}\,^{i}\,\xi_{\hat{s}}\,^{j}$, where
$\eta_{\hat {r}\hat{s}} = \eta^{\hat{r}\hat{s}} = diag(-1,1,1)$. In
terms of the cotriad the metric on $^{3}\!B$ is given by $\gamma_{ij} =
\eta_{\hat{r}\hat{s}}\,\xi^{\hat{r}}\,_{i}\,\xi^{\hat{s}}\,_{j}$ , and
$(\xi)$ represents $det|\xi^{\hat{r}}\,_{j}| = \sqrt{-\gamma}$.  At
this point the {\em densitized triad} $(\xi)\xi_{\hat{r}}\,^{j} =
\tilde{\xi}_{\hat{r}}\,^{j}$ is chosen as the fundamental variable, and
$\gamma_{ij}$ is considered to be a secondary or  derived quantity. As
a result of this shift in emphasis, the subtraction term in
(\ref{BYaction2}) is now considered to be a functional solely of the
densitized triad. Employing the identity

\begin{equation}
\frac{\partial\gamma_{ij}}{\partial\tilde{\xi}_{\hat{r}}\,^{k}} =
(\xi)^{-1}\left(\xi^{\hat{r}}\,_{k}\gamma_{ij} -
\xi^{\hat{r}}\,_{i}\gamma_{kj} -
\xi^{\hat{r}}\,_{j}\gamma_{ki}\right)\, , \end{equation} one finds that
the momentum conjugate to $\tilde{\xi}^{\hat{r}}\,_{j}$ with respect to
$^{3}\!B$ is

\begin{equation} \Pi^{\hat{r}}\,_{j} =
\frac{\partial\gamma_{ik}}{\partial\tilde{\xi}_{\hat{r}}\,^{j}}\,\pi^{ik}
= -\frac{1}{\kappa}\Theta^{i}\,_{j}\,\xi^{\hat{r}}\,_{i} =
-\frac{1}{\kappa}\Theta^{\hat{r}}\,_{j} . \label{bdmomentum2}
\end{equation} In the above formula $\Theta^{\hat{r}}\,_{j}$ is
referred to as the hybrid extrinsic curvature since it has one
coordinate and one orthonormal index.  The extrinsic curvature
$\Theta_{ij}$ is symmetric, however, $\Pi_{ij} =
\Pi^{\hat{r}}\,_{j}\xi_{\hat{r}i}$ is not manifestly symmetric. This
necessitates the introduction of the following additional constraint
associated with the three-boundary phase space:

\begin{equation} \Phi^{\hat{r}\hat{s}} :=
2\Pi^{[\hat{r}}\,_{j}\,\eta^{\hat{s}]\hat{t}}\,\tilde{\xi}_{\hat{t}}\,^{j}\,\approx\,
0.  \label{bdrotation} \end{equation} This {\em boundary rotation
constraint} is set weakly equal to zero, and the factor of $2$ is
chosen to make a correspondence with the analogous constraint
associated with the triad formalism for $\Sigma$.\cite{Brown} For the
remainder of this discussion the symbol $\approx$ is taken to mean
modulo the constraint (\ref{bdrotation}). With this convention one
writes

\begin{equation} \Pi_{ij}\,\approx\, - \frac{1}{\kappa}\Theta_{ij}\, .
\end{equation}

The triad is assumed to obey the {\em time gauge condition}. This
restriction requires that when the time leg of the triad $\xi_{\bot}$
is restricted to $B$, a generic leaf of the foliation, it must coincide
with the normal vector field $u$. Subject to this requirement the frame
and coframe are expressed {\em locally} as

\begin{eqnarray} \xi_{\bot} = \frac{1}{N}\left(\frac{\partial}{\partial
t} - V^{a}\frac{\partial}{\partial x^{a}}\right) &  &\xi_{\hat a} =
\beta_{\hat{a}}\,^{b}\frac{\partial}{\partial x^{b}} \nonumber \\
\xi^{\bot} = Ndt\hspace{2.55cm} &  & \xi^{\hat a} =
\beta^{\hat{a}}\,_{b}\left( dx^{b} + V^{b}dt\right) . \label{bdtriad}
\end{eqnarray} In the above relations $\{\beta_{\hat a}\mid\hat{a} =
\hat{1}, \hat{2}\}$ and $\{\beta^{\hat a}\mid\hat{a} = \hat{1},
\hat{2}\}$ are respectively a collection of local orthonormal frames
and a collection of local orthonormal coframes (each parameterized by
t) on the leaves $B$ of the foliated $^{3}\!B$.  An orthonormal frame
on $B$ is referred to as a {\em dyad}. The two-metric on a generic
slice $B$ is related to the coframe or codyad on the same slice by
$\sigma_{ab} =
\delta_{\hat{a}\hat{b}}\,\beta^{\hat{a}}\,_{a}\,\beta^{\hat{b}}\,_{b}$
where $\delta_{\hat{a}\hat{b}} = \delta^{\hat{a}\hat{b}} = diag(1,1)$.
On $B$ it is also true that $\delta_{\hat{a}\hat{b}} = \sigma_{ab}\,
\beta_{\hat{a}}\,^{a}\,\beta_{\hat{b}}\,^{b}$ and $\sqrt{\sigma} =
(\beta) = det|\beta^{\hat{a}}\,_{b}|$.  The densitized triad can now be
broken into three pieces: $N, V^{a},$ and
$\tilde{\beta}_{\hat{a}}\,^{b} = (\beta )\beta_{\hat{a}}\,^{b}$. {\em
One should note that} $\tilde{\beta}_{\hat{a}}\,^{b}$ {\em is a
two-density and not a three-density like}
$\tilde{\xi}_{\hat{r}}\,^{j}$. The following identities are easily
verified:

\begin{eqnarray}
  &  & \frac{\partial\tilde{\xi}_{\hat{r}}\,^{j}}{\partial N} =
\sqrt{\sigma}\,\beta_{\hat{a}}\,^{b}\,\delta^{\hat{a}}_{\hat{r}}\,\delta^{j}_{b}
\nonumber \\
  &  &  \frac{\partial\tilde{\xi}_{\hat{r}}\,^{j}}{\partial V^{b}} =
-\,\sqrt{\sigma}\,\delta^{\bot}_{\hat{r}}\,\delta^{j}_{b}
\label{identities1} \\
  &  &

\frac{\partial\tilde{\xi}_{\hat{r}}\,^{j}}{\partial\tilde{\beta}_{\hat{a}}\,^{b}}
= \beta^{\hat{a}}\,_{b}\,\delta^{\bot}_{\hat{r}}\,\delta^{j}_{0} -
V^{c}\,\beta^{\hat{a}}\,_{b}\,\delta^{\bot}_{\hat{r}}\,\delta^{j}_{c} +
N\,\delta^{\hat{a}}_{\hat{r}}\,\delta^{j}_{b}\, .  \nonumber
\end{eqnarray} If the subtraction term $S^{0}$ in (\ref{BYaction2}) is
taken to be zero, then these identities allow one to demonstrate
swiftly that the Brown-York quasilocal energy and momentum densities
are given by

\begin{eqnarray}
  &  &  \varepsilon |_{_{0}}\: =
-\,\beta_{\hat{a}}\,^{b}\,\Pi^{\hat{a}}\,_{b}\,\approx\,\frac{1}{\kappa}\,k
\label{epzero} \\
  &  &  j_{b} |_{_{0}}\: = -\,\Pi^{\bot}\,_{b}\,\approx\, -
\frac{2}{\sqrt{h}}\,\sigma_{b\underline{i}}\,n_{\underline{j}}\,p^{\underline{ij}}\,
.  \label{jayzero} \end{eqnarray} The symbol $|_{_{0}}$ is used to
indicate the choice $S^{0} =0$.  In the above relations $k$ is the
trace of the extrinsic curvature $k_{ab}$ of $B$ as a two-surface
embedded in $\Sigma$. Also, the mixed component two-metric in
(\ref{jayzero}) may be expressed as

\begin{equation} \sigma_{b\underline{i}} = \frac{\partial
x^{c}}{\partial x^{\underline{i}}}\,\sigma_{bc}\, .  \end{equation} The
final expressions in (\ref{epzero}) and (\ref{jayzero}) are obtained by
the use of a Brown-York result.\cite{BY} They have found the following
expression for the three-boundary extrinsic curvature:

\begin{equation} \Theta_{ij} = k_{ij} +
u_{i}u_{j}n_{\underline{k}}a^{\underline{k}} +
2\sigma^{\underline{k}}\,_{(i}u_{j)}n^{\underline{l}}K_{\underline{kl}}\,
, \end{equation} where $a = \nabla_{u}\, u$ is the spacetime covariant
acceleration of $u$.  For the above relation one should also note that
$u^{i} = \xi_{\bot}\,^{i}$ and that the two-boundary extrinsic
curvature, though it carries three-boundary indices, is a spatial
tensor because $k_{ij}\, u^{j} = 0$.

\section{Boundary Ashtekar variables}

\subsection{Construction and properties}

The introduction of the boundary Ashtekar variables is achieved by
appending to the action (\ref{BYaction1}) the following pure imaginary
subtraction term:  \begin{equation} S^{0} =
\frac{i}{2\kappa}\int_{^{3}\!B}d^{3}x\,
\epsilon^{\hat{r}\hat{s}\hat{t}}\,\tau_{\hat{s}\hat{t}
j}\tilde{\xi}_{\hat{r}}\,^{j}\, .  \label{bdsubterm} \end{equation} The
$\tau^{\hat{r}}\,_{\hat{s}\hat{t}} = \tau^{\hat{r}}\,_{\hat{s}
i}\,\xi_{\hat{t}}\,^{i}$ are the connection coefficients determined by
the time gauge triad. These are given explicitly by

\begin{eqnarray}
  &  & \tau^{\bot}\,_{\hat{c}\bot} = \beta_{\hat{c}}\left[\log N\right]
\nonumber \\
  &  & \tau^{\bot}\,_{\hat{a}\hat{c}} = -\frac{1}{N}\left(
\sigma_{bd}\,\beta_{(\hat{a}}\,^{b}\,\dot{\beta}_{\hat{c})}\,^{d} +\,
^{2}\overline{\nabla}_{(\hat{c}}V_{\hat{a})}\right)
\label{bdconnection} \\
  &  & \tau^{\hat{1}}\,_{\hat{2}\hat{a}} =
  \beta_{\hat{a}}\,^{a}\beta^{\hat{1}}\,_{b}\, \left(
^{2}\nabla_{a}\,\beta_{\hat{2}}\,^{b}\right) \nonumber \\
  &  & \tau_{\hat{1}\hat{2}\bot} =
\frac{1}{N}\left(\sigma_{bd}\,\beta_{[\hat
{1}}\,^{b}\,\dot{\beta}_{\hat{2} ]}\,^{d} +\,
^{2}\overline{\nabla}_{[\hat {2}}V_{\hat{1} ]} -
V^{\hat{c}}\,\tau_{\hat{1}\hat{2}\hat{c}}\right)\, , \nonumber
\end{eqnarray} where the dot denotes partial time differentiation. The
choice of this particular subtraction term may appear {\em ad hoc},
however, its introduction is justified by a careful treatment of the
boundary terms of an appropriate complex tetrad action principle for
general relativity.  The analysis of this complex tetrad action
principle is found in the the last section. From the form of these time
gauge connection coefficients one can verify that the subtraction term
(\ref{bdsubterm}) is functionally linear in both $N$ and $V^{a}$. These
two conditions imply that the alterations of $\varepsilon |_{_{0}}$ and
$j_{b} |_{_{0}}$ which arise from appending (\ref{bdsubterm}) to the
action (\ref{BYaction1}) do not spoil the requirement that these
quantities depend only on the Cauchy data of $\Sigma$.\cite{BY}

With the full action (\ref{BYaction2}) the three-boundary momentum
conjugate to the densitized triad is given by

\begin{equation} \Pi_{_{NEW}}\,^{\hat{r}}\,_{j} = \Pi^{\hat{r}}\,_{j} -
\frac{i}{2\kappa}\,\epsilon^{\hat{r}\hat{s}\hat{t}}\,\tau_{\hat{s}\hat{t}j}\,
, \label{bdmomentum3} \end{equation} because the term (\ref{bdsubterm})
obeys a remarkable identity which is analogous to an identity often
employed in the construction of the $\Sigma$ Ashtekar
variables.\cite{tetradgrav} Namely,

\begin{equation} \frac{\delta S^{0}}{\delta\tilde{\xi}_{\hat{r}}\,^{j}}
=
\frac{i}{2\kappa}\,\epsilon^{\hat{r}\hat{s}\hat{t}}\,\tau_{\hat{s}\hat{t}j}\,
. \label{identity2} \end{equation} The boundary Ashtekar connection is
defined by

\begin{eqnarray} {\cal A}^{\hat{r}}\,_{j}\,:=
i\Pi_{_{NEW}}\,^{\hat{r}}\,_{j} & = &
\frac{1}{\kappa}\,\tau^{\hat{r}}\,_{j} + i\Pi^{\hat{r}}\,_{j}
\label{bdAshconnection} \\
  & \approx & \frac{1}{\kappa}\left( \tau^{\hat{r}}\,_{j} -
i\Theta^{\hat{r}}\,_{j}\right)\, , \nonumber \end{eqnarray} where
$\tau^{\hat{r}}\,_{j} =
\frac{1}{2}\epsilon^{\hat{r}\hat{s}\hat{t}}\,\tau_{\hat{s}\hat{t}j}$.
Modulo the boundary rotation constraint (\ref{bdrotation}), the
boundary Ashtekar connection (\ref{bdAshconnection}) is the Sen
connection \cite{Ashtekar1,Sen} for the three-boundary $^{3}\!B$. The
transformation of the $^{3}\! B$ phase space coordinates obtained by
appending (\ref{bdsubterm}) to the action is of the form $(q^{j},
p_{j}) \rightarrow (q^{j}, ip_{j} + \partial G/\partial q^{j})$, where
$G\left[ q^{j}\right]$ is a q-dependent generating function. The
boundary Ashtekar variable ${\cal A}^{\hat{r}}\,_{j}$ is the pull back
of a complexified $SO(2,1)$ connection on the bundle of orthonormal
frames over $^{3}\!B$. The associated derivative operator is defined by
its action on vector fields $v^{\hat{r}}$

\begin{equation} \hat{{\cal D}}_{j}\, v^{\hat{r}} =   \partial_{j}\,
v^{\hat{r}} + \kappa\,\epsilon^{\hat{r}\hat{s}}\,_{\hat{t}}\,{\cal
A}_{\hat{s}j}\,v^{\hat{t}}\, .  \end{equation} The boundary rotation
constraint (\ref{bdrotation}) can be expressed in a ``Gauss law" form
with this derivative operator. Indeed, the identity $\overline{{\cal
D}}_{j}\tilde{\xi}_{\hat{k}}\,^{j} = 0$ justifies the following
expression:  \begin{equation} \hat{{\cal
D}}_{j}\tilde{\xi}_{\hat{r}}\,^{j} =
\frac{i}{2}\kappa\,\epsilon_{\hat{r}\hat{s}\hat{t}}\,\Phi^{\hat{s}\hat{t}}\,
.\label{bdGauss} \end{equation} Since the epsilon symbol is invertible,
the {\em boundary Gauss constraint} (\ref{bdGauss}) is fully equivalent
to the boundary rotation constraint (\ref{bdrotation}).

Because $^{3}\!B$ is a submanifold of the metric space $M$, its
geometry must obey the Gauss-Codacci embedding conditions. These are
integrability criteria relating the spacetime Riemann tensor
$\Re_{\alpha\beta\mu\nu}\left[ g\right]$ to the $^{3}\!B$ Riemann
tensor ${\cal R}_{ijkl}\left[ \gamma\right]$ and extrinsic curvature
tensor $\Theta_{ij}$.\cite{York} Further, if $M$ is an Einstein
spacetime, then the vacuum field equations $\Re_{\alpha\beta} = 0$
hold. The Gauss-Codacci conditions are used to express the projections
$n^{\alpha}\, n^{\beta}\,\Re_{\alpha\beta}$ and
$\gamma^{\alpha}_{\mu}\, n^{\beta}\,\Re_{\alpha\beta}$ in terms of the
triple $\left(^{3}\!B, \gamma_{ij},\Theta_{ij}\right)$. The $^{3}\!B$
metric is given in spacetime coordinates by $\gamma_{\mu\nu} =
g_{\mu\nu} - n_{\mu}\, n_{\nu}$. Apart from density factors, these
projections are respectively the {\em boundary scalar constraint} and
{\em boundary vector constraint}. The curvature of ${\cal
A}^{\hat{r}}\,_{j}$,

\begin{eqnarray} {\cal F}^{\hat{r}}\,_{jk} & = & 2\partial_{[j}{\cal
A}^{\hat{r}}\,_{k]} + \kappa\,\epsilon^{\hat{r}\hat{s}\hat{t}}\,{\cal
A}_{\hat{s}j}\,{\cal A}_{\hat{t}k} \label{bdAshcurvature}\\
  & = &  \frac{1}{\kappa}{\cal R}^{\hat{r}}\,_{jk} -
\kappa\,\epsilon^{\hat{r}\hat{s}\hat{t}}\,\Pi_{\hat{s}j}\,\Pi_{\hat{t}k}
+ 2i\,\overline{{\cal D}}_{[j}\,\Pi^{\hat{r}}\,_{k]}\, , \nonumber
\end{eqnarray} can be used to express these constraints compactly. In
the above formula (\ref{bdAshcurvature}) one should note that

\begin{equation} {\cal R}^{\hat{r}}\,_{jk} =
2\partial_{[j}\,\tau^{\hat{r}}\,_{k]} +
\epsilon^{\hat{r}\hat{s}\hat{t}}\,\tau_{\hat{s}j}\,\tau_{\hat{t}k}\, ,
\end{equation} where the relation of this Yang-Mills-type curvature to
the mixed components of the Riemann tensor on $^{3}\!B$ is given by
${\cal R}^{\hat{r}}\,_{\hat{t}jk} =
\epsilon^{\hat{r}}\,_{\hat{s}\hat{t}}\,{\cal R}^{\hat{s}}\,_{jk}$. In
an orthonormal frame the Riemann tensor on $^{3}\!B$ takes its values
in the Lie algebra of the matrix group $SO(2,1)$. It is evident from
the considerations above that the theory employs the adjoint or vector
representation for the generators $\left\{ T_{\hat{r}} | \hat{r} =
\bot,\hat{1},\hat{2}\right\}$. In other words, the Lie algebra valued
forms on $^{3}\!B$ are being contracted on the structure constants
$\epsilon^{\hat{r}}\,_{\hat{s}\hat{t}} = \left(
T_{\hat{s}}\right)^{\hat{r}}\,_{\hat{t}}$ of the $so(2,1)$ Lie
algebra.  Manipulation of the boundary Ashtekar curvature
(\ref{bdAshcurvature}) yields the following elegant expressions for the
boundary scalar and boundary momentum constraints:

\begin{eqnarray} \frac{1}{2(\xi )}\,\epsilon^{\hat{r}\hat{s}\hat{t}}\,
\tilde{\xi}_{\hat{r}}\,^{i}\,\tilde{\xi}_{\hat{s}}\,^{j}\,{\cal
F}_{\hat{t}ij} & = & \frac{\kappa}{2}\sqrt{-\gamma}\left\{
(\Pi^{j}\,_{j})^{2} - \Pi^{i}\,_{j}\,\Pi^{j}\,_{i}\right\} \nonumber \\
 &  & - \frac{\sqrt{-\gamma}}{2\kappa}\,{\cal R} -\,
i\,\sqrt{-\gamma}\,\epsilon^{ijk}\,{\cal D}_{i}\,\Pi_{jk}
\label{bdscalar} \\
   & \approx & \frac{\sqrt{-\gamma}}{2\kappa}\left\{
   (\Theta^{i}\,_{i})^{2} - \Theta^{i}\,_{j}\,\Theta^{j}\,_{i} - {\cal
R}\right\}\, ,  \nonumber \end{eqnarray}

\begin{eqnarray} i\,\tilde{\xi}_{\hat{r}}\,^{i}\,{\cal
F}^{\hat{r}}\,_{ij} & = &
 - \frac{i}{2\kappa}\sqrt{-\gamma}\,\epsilon^{lmp}\,{\cal R}_{j[lmp]}
 \nonumber \\
  &  &  + \,i\kappa\,\sqrt{-\gamma}\,\epsilon^{lmp}\,\Pi_{lm}\,\Pi_{pj}
  - 2 \,\sqrt{-\gamma}\,{\cal D}_{[i}\,\Pi^{i}\,_{j]}  \label{bdvector}
\\
 & \approx & \frac{2}{\kappa}\,\sqrt{-\gamma}\,{\cal
 D}_{[i}\,\Theta^{i}\,_{j]}\, . \nonumber \end{eqnarray} In
(\ref{bdscalar}) the $\epsilon^{ijk} =
\epsilon^{\hat{r}\hat{s}\hat{t}}\,\xi_{\hat{r}}\,^{i}\,\xi_{\hat{s}}\,^{j}\,\xi_{\hat{t}}\,^{k}$
 are the coordinate components of the Levi-Civita tensor, and ${\cal
 R}$ denotes the Ricci scalar of $^{3}\!B$. The last step in
(\ref{bdvector}) appeals to the algebraic Bianchi identity for the
Riemann tensor.\cite{DiffGeom} It is interesting to note that the
relative signs of the quadratic extrinsic curvature terms in
(\ref{bdscalar}) are like those found for a hypersurface in a Euclidean
spacetime. This being true, one may wonder why the factor of $i$
multiplying $\Pi^{\hat{r}}\,_{j}$ is necessary in the definition of the
boundary Ashtekar connection, since the Ashtekar connection for the
Euclidean case is manifestly real. However, the $i$ is crucial, because
$^{3}\!B$ does have a time direction and thus there is an extra minus
sign hidden in its Levi-Civita tensor used to form (\ref{bdscalar}).
Since $^{3}\!B$ is a Lorentz manifold, there are also extra minus signs
buried in the Ricci scalar because it is built from a metric with
signature $(-1,1,1)$. However, these signs have no effect on the form
of (\ref{bdscalar}).

Some mention of reality conditions for these boundary variables is in
order.  The following nonpolynomial form for these conditions suffices
for this work:

\begin{eqnarray} \left( \tilde{\xi}_{\hat{r}}\,^{j}\right)^{*} & = &
\tilde{\xi}_{\hat{r}}\,^{j} \nonumber \\
 \left( {\cal A}^{\hat{r}}\,_{j} - \frac{1}{\kappa}\,
 \tau^{\hat{r}}\,_{j}\right)^{*} & = & -  \left( {\cal
 A}^{\hat{r}}\,_{j} - \frac{1}{\kappa}\, \tau^{\hat{r}}\,_{j}\right)\,
 , \end{eqnarray} where $*$ denotes complex conjugation.

\subsection{Relation to Brown-York quasilocal expressions}

In the previous expressions for the quasilocal energy surface density
(\ref{epzero}) and the quasilocal momentum surface density
(\ref{jayzero}) it was assumed that the subtraction term was zero.
After appending the subtraction term (\ref{bdsubterm}) to the action
(\ref{BYaction1}), an examination of the definitions for $\varepsilon$
(\ref{ep}) and $j_{b}$ (\ref{jay}) shows that

\begin{eqnarray} \varepsilon & = & i\,\beta_{\hat{a}}\,^{b}\,{\cal
A}^{\hat{a}}\,_{b} \nonumber \\  & = &
-\beta_{\hat{a}}\,^{b}\,\Pi^{\hat{a}}\,_{b} +
\frac{i}{\kappa}\,\epsilon^{\bot\hat{a}\hat{c}}\,\tau_{\bot\hat{a}\hat{c}}
\label{bdAshep} \\
 & = & \varepsilon |_{_{0}}\nonumber \\     &    &    \nonumber \\
j_{b} & = & i\,{\cal A}^{\bot}\,_{b}  \nonumber \\ & = &
-\Pi^{\bot}\,_{b} - \frac{i}{\kappa}\,\tau_{\hat{1}\hat{2}b}
\label{bdAshjay}  \\
     & = & j_{b} |_{_{0}} - \frac{i}{\kappa}\,\tau_{\hat{1}\hat{2}b}\,
     . \nonumber \end{eqnarray} The last equality in (\ref{bdAshep})
holds because $\tau_{\bot\hat{a}\hat{c}} =
\tau_{\bot(\hat{a}\hat{c})}$. Provided that the time gauge condition is
enforced, as it is here, the expression for the Brown-York quasilocal
energy density $\varepsilon$ is unchanged relative to the $S^{0} = 0$
case by this choice of subtraction term. The momentum density, however,
picks up an imaginary piece. This term is proportional to the
connection form on $B$, and hence is related to the freedom to perform
local dyad rotations.

Expanding the expression $\tilde{\xi}_{\hat{r}}\,^{a}{\cal
A}^{\hat{r}}\,_{a}$ which has a ``qp" form, one verifies that

\begin{eqnarray} \tilde{\xi}_{\hat{r}}\,^{b}{\cal A}^{\hat{r}}\,_{b} &
= & \sqrt{-\gamma}\,\xi_{\hat{r}}\,^{b}{\cal A}^{\hat{r}}\,_{b}
\nonumber \\
  & = &  N\sqrt{\sigma}\,\xi_{\bot}\,^{b}{\cal A}^{\bot}\,_{b} +
N\sqrt{\sigma}\,\beta_{\hat{a}}\,^{b}{\cal A}^{\hat{a}}\,_{b}
\label{bdHamdensity}\\
  & = &  - i\sqrt{\sigma}\left( -V^{b}j_{b} + N\varepsilon\right)\, .
  \nonumber \end{eqnarray} It should be noted that in the proceeding
double sum the coordinate indices run only over $(1,2)$. The previous
formula is the main result of this work.  Apart from a factor of $- i$
the above expression is the integrand of the boundary piece of the
Hamiltonian that Brown and York find via a canonical analysis of
(\ref{BYaction2}).\cite{BY} Therefore, the now complex surface term in
the Hamiltonian can be written compactly as

\begin{equation} H_{\partial\Sigma} =
i\int_{B}d^{2}x\,\tilde{\xi}_{\hat{r}}\,^{a}{\cal A}^{\hat{r}}\,_{a}\,
.  \label{bdHamiltonian} \end{equation}

The close relation between the Brown-York expressions and the boundary
Ashtekar variables is made manifest by selecting the time gauge. One
could consider subtraction terms of the same form as (\ref{bdsubterm})
but built from arbitrary triads and their associated connection forms.
Such a more general subtraction term would also allow the introduction
of a boundary Ashtekar connection. However, in general such a
subtraction term would not be functionally linear in $N$ and $\vec{V}$,
and therefore the resulting expressions for the quasilocal energy and
momentum densities would not depend solely on slice Cauchy data. In
other words, these more general subtraction terms would mix the slice
dynamical variables (Cauchy data) with essentially ``gauge" variables.
It does not seem that anything physically significant would be gained
by allowing such generality.

\section{Sparling forms}

The Brown-York expressions for the $S^{0} = 0$ quasilocal energy and
momentum surface densities, relations (\ref{epzero}) and
(\ref{jayzero}), are intimately related to certain frame-dependent
differential forms first introduced by Sparling.  This section presents
the construction of these forms, a review of their properties, and
their relation to the original Brown-York and boundary Ashtekar
variables.

\subsection{Construction and properties}

Before the introduction of the Sparling forms can be made, it is first
necessary to consider on the spacetime $M$ an orthonormal frame
(technically pseudo-orthonormal) or {\em tetrad} $\{ e_{\hat{\alpha}} |
\hat{\alpha} = \bot, \hat{1}, \hat{2}, \hat{3}\} $ and its associated
coframe or cotetrad $\{ e^{\hat{\alpha}} | \hat{\alpha} = \bot,
\hat{1}, \hat{2}, \hat{3}\}$.  A spacetime manifold $M$ which is
topologically $R\times \Sigma$ admits a globally defined orthonormal
frame, if $\Sigma$ is an orientable three-space.\cite{DiffGeom}
However, there will be need to examine certain frames which cannot in
general be globally defined.  In a coordinate system the tetrad and
cotetrad may be expanded respectively as $ e_{\hat{\alpha}} =
e_{\hat{\alpha}}\,^{\mu}\,\partial_{\mu}$ and $e^{\hat{\alpha}} =
e^{\hat{\alpha}}\,_{\mu}\,dx^{\mu}$.  The coordinate components of the
Lorentz metric on $M$ are given by $g_{\mu\sigma} =
\eta_{\hat{\alpha}\hat{\beta}}e^{\hat{\alpha}}\,_{\mu}e^{\hat{\beta}}\,_{\sigma}$
and $e_{\hat{\alpha}}\,^{\mu}e_{\hat{\beta}\mu} =
\eta_{\hat{\alpha}\hat{\beta}}$ , where $\eta_{\hat{\alpha}\hat{\beta}}
= \eta^{\hat{\alpha}\hat{\beta}} = diag(-1,1,1,1)$.

For such a tetrad the connection one-forms
$\omega_{\hat{\alpha}\hat{\beta}} =
\omega_{\hat{\alpha}\hat{\beta}\hat{\gamma}}e^{\hat{\gamma}} =
\omega_{\hat{\alpha}\hat{\beta}\mu}dx^{\mu}$ which specify the
Levi-Civita connection on the frame bundle of $M$ are characterized by
the following two requirements:\cite{DiffGeom} (1) the metric
compatibility condition

\begin{equation} \omega_{\hat{\alpha}\hat{\beta}}
=\omega_{[\hat{\alpha}\hat{\beta}]}\, , \label{metriccompatibility}
\end{equation} and (2) the vanishing of the torsion two-form which
casts the first Cartan structure equation into the form

\begin{equation} de^{\hat{\alpha}} +
\,\omega^{\hat{\alpha}}\,_{\hat{\beta}} \wedge e^{\hat{\beta}} = 0\, .
\label{notorsion} \end{equation} The curvature two-form is defined by
the second Cartan structure equation

\begin{equation} \Omega^{\hat{\alpha}}\,_{\hat{\beta}} =
d\omega^{\hat{\alpha}}\,_{\hat{\beta}} +
\omega^{\hat{\alpha}}\,_{\hat{\gamma}} \wedge
\omega^{\hat{\gamma}}\,_{\hat{\beta}}\, , \end{equation} and the
orthonormal components of the spacetime Riemann tensor are related to
the curvature form by the relation

\begin{equation} \Omega^{\hat{\alpha}}\,_{\hat{\beta}} =
\frac{1}{2}\Re^{\hat{\alpha}}\,_{\hat{\beta}\hat{\gamma}\hat{\delta}}
e^{\hat{\gamma}} \wedge e^{\hat{\delta}}.  \end{equation}

{}From the cotetrad and the connection forms one can construct the
Sparling
two-forms and three-forms

\begin{equation} \sigma_{\hat{\alpha}} =
-\,\frac{1}{2}\,\omega^{\hat{\beta}\hat{\gamma}} \wedge
e^{\ast}_{\hat{\alpha}\hat{\beta}\hat{\gamma}}   \label{sigSparling}
\end{equation}

\begin{equation} \tau_{\hat{\alpha}} =
\frac{1}{2}\left(\omega_{\hat{\alpha}}\,^{\hat{\beta}} \wedge
\omega^{\hat{\gamma}\hat{\delta}} \wedge
e^{\ast}_{\hat{\beta}\hat{\gamma}\hat{\delta}} -
\omega^{\hat{\beta}}\,_{\hat{\delta}} \wedge
\omega^{\hat{\delta}\hat{\gamma}} \wedge
e^{\ast}_{\hat{\alpha}\hat{\beta}\hat{\gamma}}\right)\, ,
\label{tauSparling} \end{equation} where
$e^{\ast}_{\hat{\alpha}\hat{\beta}\hat{\gamma}} =
\epsilon_{\hat{\alpha}\hat{\beta}\hat{\gamma}\hat{\delta}}e^{\hat{\delta}}$.

Several properties of the Sparling forms should be noted. Since the
connection coefficients transform inhomogeneously under tetrad
transformations, the Sparling forms defined above are frame dependent.
However, it is possible to construct canonically defined analogs of
these forms on the orthonormal frame bundle over $M$. Pull backs of
these canonically defined forms with respect to sections of the
orthonormal frame bundle yield the various frame-dependent Sparling
forms on $M$.\cite{Sparling}  The fundamental relation obeyed by the
Sparling forms is now presented. Employing the structure equations and
the fact that $G_{\hat{\alpha}\hat{\beta}}$, the tetrad version of the
Einstein tensor, is the contracted double dual Riemann tensor
\cite{MTW}, one can demonstrate that

\begin{equation} d\sigma_{\hat{\alpha}} = \tau_{\hat{\alpha}} +
G_{\hat{\alpha}}\,^{\hat{\beta}}e^{\ast}_{\hat{\beta}}\, ,
\label{Sparlingrelation} \end{equation} where $e^{\ast}_{\hat{\alpha}}
=
(3!)^{-1}\,\epsilon_{\hat{\alpha}\hat{\beta}\hat{\gamma}\hat{\delta}}e^{\hat{\beta}}
\wedge e^{\hat{\gamma}} \wedge e^{\hat{\delta}}$. The form of the above
equation suggests that $\tau_{\hat{\alpha}}$ may be interpreted as a
frame-dependent energy-momentum density with the corresponding
superpotential given by $\sigma_{\hat{\alpha}}$.\cite{Goldberg} For the
case of this discussion the form of (\ref{Sparlingrelation}) is even
simpler since $G_{\hat{\alpha}\hat{\beta}} = 0$ for vacuum spacetimes.
In fact, for the vacuum theory the fundamental relation
(\ref{Sparlingrelation}) reduces Einstein's equations to an
integrability theorem: $M$ is Ricci flat if and only if
$d\sigma_{\hat{\alpha}} = \tau_{\hat{\alpha}}$.

\subsection{Link with $^{3}\! B$ variables}

Since $\sigma_{\hat{\alpha}}$ acts like a frame-dependent
superpotential in (\ref{Sparlingrelation}), an examination of its form
for a natural choice of frame is merited. In the remainder of this
section a natural choice of frame is constructed, and the Brown-York
expressions for $\varepsilon |_{_{0}}$ (\ref{epzero}) and $j_{b}
|_{_{0}}$ (\ref{jayzero}) are written in terms of the connection
coefficients associated with this distinguished frame. Finally, the
relation between $\sigma_{\hat{\alpha}}$ and the Brown-York formalism,
cast in this connection coefficient language, is revealed.

Though the frame which will be considered is quite natural, its
construction requires some effort. For some $X \subset B$, an open
subset of a generic leaf of the foliated $^{3}\!B$, it is assumed that
there exists a dyad $\left\{\beta_{\hat{a}} | \hat{a} = \hat{1},
\hat{2}\right\}$ globally defined on $X$. There is also need to
consider $Y \subset \Sigma$, an open set which contains $B$ and
therefore $X$. It is assumed that there exists a function $r: Y
\rightarrow {\cal I}$, where ${\cal I}$ is a connected closed subset of
the real line. The function $r$ provides a local ``radial foliation" of
$\Sigma$ in the region surrounding $B$. The leaves of this local
foliation have the topology of $B$. The foliation associated with the
function $r$ may not be extendable to one which fills the whole
spacelike slice $\Sigma$. The construction of the distinguished frame
proceeds as follows: first employing the radial foliation, one
constructs a triad $\left\{ E_{\hat{s }} |\, \hat{s} = \,\vdash,
\hat{1}, \hat{2}\right\}$ and its associated cotriad on $\Sigma$ from
the dyad and codyad on $B$; and then one employs the hypersurface
foliation $\Sigma$ to construct a tetrad and cotetrad from this slice
triad and cotriad. This prescription gives the following tetrad and
cotetrad on a suitably small spacetime neighborhood of $X$:

\begin{eqnarray} e_{\bot} = u = \xi_{\bot} =
\frac{1}{N}\left(\frac{\partial}{\partial t} -
V^{\hat{a}}\beta_{\hat{a}} - V^{\vdash}E_{\vdash}\right) &  & e^{\bot}
= \xi^{\bot} = Ndt \nonumber \\ e_{\vdash} =\,^{_{2}}n = E_{\vdash} =
\frac{1}{M}\left(\frac{\partial}{\partial r} -
W^{\hat{a}}\beta_{\hat{a}}\right)  &  & e^{\vdash} = E^{\vdash} +
V^{\vdash}dt \nonumber  \\
  &  & e^{\vdash} = M dr + V^{\vdash}dt\hspace{1.75cm}
  \label{distinguishedframe} \end{eqnarray}

\begin{eqnarray} \hspace{3.25cm}e_{\hat{a}} = \xi_{\hat{a}} =
E_{\hat{a}} = \beta_{\hat{a}} &  &  e^{\hat{a}} = \beta^{\hat{a}} +
V^{\hat{a}}dt + W^{\hat{a}}dr \nonumber \\
  &  & e^{\hat{a}} = \xi^{\hat{a}} + W^{\hat{a}}dr  \nonumber \\ &  &
  e^{\hat{a}} = E^{\hat{a}} + V^{\hat{a}}dt  \nonumber \end{eqnarray}
For the above batch $M$ and $W^{\hat{a}}$ are respectively the
kinematical ``lapse" and ``shift" associated with the local radial
foliation. One should recall that in this analysis the crucial
assumption is made that the leaves of the hypersurface foliation
$\Sigma$ are orthogonal to $^{3}\!B$. From the canonical perspective
this assumption requires that the component of $\vec{V}$ normal to $B$
must vanish \cite{BY}

\begin{equation} \left.\vec{V} \cdot e_{\vdash} \right|_{^{3}\! B} =
0\, .\label{distinguishedgauge} \end{equation} Therefore, one may set
$V^{\vdash} = 0$ in the above collection for the remainder of this
section. It follows that $e_{\vdash} =\, ^{_{2}}n = n$ is also the
normal of $^{3}\! B$ in $M$.

At this point, in order to make a comparison with the Brown-York and
boundary Ashtekar quasilocal expressions, an auxiliary assumption is
made. {\em It is assumed that the spacetime metric on the region
surrounding $X$ determined by the distinguished tetrad}
(\ref{distinguishedframe}) {\em is Ricci flat}. In other words, the
metric

\begin{equation} g = -\, e^{\bot} \otimes e^{\bot} + e^{\vdash} \otimes
e^{\vdash} + e^{\hat{1}} \otimes e^{\hat{1}} + e^{\hat{2}} \otimes
e^{\hat{2}} \end{equation} is a solution of the vacuum Einstein
equations. This requirement is enforced for the remainder of this (4.2)
subsection, and hence most of the superfluous $c\ell$'s which should be
present in the following formulas will be suppressed. Since this
auxiliary assumption is made, the quasilocal expressions (\ref{epzero})
and (\ref{jayzero}) may be expressed in terms of the connection
coefficients determined by (\ref{distinguishedframe}). The extrinsic
curvature of $B$ as embedded in $\Sigma$ is given by

\begin{equation} k_{ij} = - \sigma_{i}\,^{k}\, D_{k}n_{j}\, ,
\end{equation} where the two-metric on $B$ is given in $\Sigma$
coordinates by $\sigma_{ij} = h_{ij} - n_{i}n_{j}$.  Therefore, for an
arbitrary orthonormal triad $\left\{ E_{\hat{r}} = e_{\hat{r}} |
\hat{r} = \hat{1},\hat{2},\hat{3}\right\}$ on $\Sigma$ the $S^{0} = 0$
quasilocal energy density (\ref{epzero}) is expressed by

\begin{equation} \varepsilon |_{_{0}} = - \frac{1}{\kappa}\left(
E_{\hat{r}}\left[ n^{\hat{r}}\right] +
n^{\hat{s}}\,\omega^{\hat{r}}\,_{\hat{s}\hat{r}}\right)\, .
\end{equation} However, when the triad on $\Sigma$ in the region near
$B$ is determined by the pull back of (\ref{distinguishedframe}), the
orthonormal indices run over $(\vdash, \hat{1}, \hat{2})$. In this case
the above equation allows one to write

\begin{equation} \varepsilon |_{_{0}} = \left. -
\frac{1}{\kappa}\,\omega^{\hat{a}}\,_{\vdash\hat{a}} \right|_{c\ell}\,
.\label{connectionep} \end{equation} The $c\ell$ notation is again
adopted to emphasize the extra condition that $g$ is Ricci flat.
Similarly, for an arbitrary triad on $\Sigma$ one can express the
$S^{0} = 0$ quasilocal momentum density (\ref{jayzero}) as

\begin{equation} j_{b} |_{_{0}} =  - \frac{1}{\kappa}\,
E^{\underline{\hat{r}}}\,_{b}\,n^{\underline{\hat{s}}}\,
\omega^{\bot}\,_{\underline{\hat{r}\hat{s}}} \, .  \end{equation} The
last line above follows from (\ref{slicemomentum1}) and the definition
for the tetrad components of the extrinsic curvature
$K^{\hat{r}\hat{s}}$ in the time gauge

\begin{equation} K^{\hat{s}}\,_{\hat{r}} = \left\langle e^{\hat{s}}\,
,\, - \overline{\nabla}_{\hat{r}}\, e_{\bot}\right\rangle =
-\omega^{\hat{s}}\,_{\bot\hat{r}}\, . \label{excurvature}
\end{equation} Therefore, referring again to the distinguished frame
(\ref{distinguishedframe}) one casts the previous expression for the
quasilocal momentum density into the form

\begin{equation} j_{b} |_{_{0}} = \left. -
\frac{1}{\kappa}\,\beta^{\hat{a}}\,_{b}\,\omega^{\bot}\,_{\vdash\hat{a}}
\right|_{c\ell}\, . \label{connectionjay} \end{equation} Again, the
${c\ell}$ notation has been reintroduced for emphasis.

The Sparling two-forms $\left\{ \sigma_{\bot}, \sigma_{\vdash},
\sigma_{\hat{a}}\right\}$  are built with the connection coefficients
$\omega^{\hat{\alpha}}\,_{\hat{\beta}\mu}$ determined by
(\ref{distinguishedframe}) using the definition (\ref{sigSparling}).
The $\sigma_{\bot}$ and $\sigma_{\hat{a}}$ two-forms are quickly
related to the expressions (\ref{connectionep}) and
(\ref{connectionjay}). The inclusion map of the spacetime region $X
\subset M$ under consideration is denoted by $s: X \rightarrow M$.
This inclusion map is now used to pull back the Sparling two-forms
$\sigma_{\bot}$ and $\sigma_{\hat{a}}$ onto $B$. For $\sigma_{\bot}$
one calculates that

\begin{eqnarray} s^{*}\left( \sigma_{\bot}\right) & = & -\frac{1}{2}\,
\omega^{\hat{\alpha}\hat{\beta}}\,_{\hat{\gamma}}\,\epsilon_{\bot
\hat{\alpha}\hat{\beta}\hat{\delta}}\,\, s^{*}\left( e^{\hat{\gamma}}
\wedge e^{\hat{\delta}}\right) \nonumber\\
  & = & -
  \frac{1}{2}\,\omega^{\hat{\alpha}\hat{\beta}}\,_{\hat{a}}\,\epsilon_{\bot
  \hat{\alpha}\hat{\beta}\hat{b}}\,\beta^{\hat{a}} \wedge
  \beta^{\hat{b}} \\ & = & -\,\kappa\, \varepsilon
 |_{_{0}}\,\sqrt{\sigma}d^{2}x\, . \nonumber \end{eqnarray} Similarly,
one can pull back $\sigma_{\hat{a}}$ to discover that

\begin{equation} s^{*}\left( \sigma_{\hat{a}}\right) =
\kappa\,\beta_{\hat{a}}\,^{b}\,j_{b} |_{_{0}}\,\sqrt{\sigma} d^{2}x\,
.  \end{equation} The previous considerations allow one to write the
following compact expression:

\begin{equation}
-\,\frac{1}{\kappa}\,s^{*}\left(e^{\hat{r}}\,_{0}\,\sigma_{\hat{r}}\right)
= \left(N\,\varepsilon |_{_{0}} - V^{b}\,j_{b}
|_{_{0}}\right)\,\sqrt{\sigma}\,d^{2}x\, , \label{SparlingHam}
\end{equation} {\em where one should note that the index} $\hat{r}$
{\em is a}  $^{3}\!B$ {\em orthonormal index and hence runs over the
values} $(\bot, \hat{a} = \hat{1},\hat{2})$.

The path between the Sparling forms and the boundary Ashtekar variables
is almost as direct. First one defines the {\em self-dual} and {\em
antiself-dual connection forms} by

\begin{equation} \omega^{(\pm)\hat{\alpha}\hat{\beta}} =
\frac{1}{2}\left(\omega^{\hat{\alpha}\hat{\beta}} \mp
\frac{i}{2}\epsilon^{\hat{\alpha}\hat{\beta}\hat{\gamma}\hat{\delta}}\omega_{\hat{\gamma}\hat{\delta}}\right).
\label{sd/asdconnection} \end{equation} These complex connection forms
suggest the following complex Sparling superpotentials:

\begin{equation} \sigma^{(\pm)}\,_{\hat{\alpha}} =
-\omega^{(\pm)\hat{\beta}\hat{\gamma}} \wedge
e^{\ast}_{\hat{\alpha}\hat{\beta}\hat{\gamma}}.
\label{sd/asdsigSparling} \end{equation} For the distinguished gauge
choice (\ref{distinguishedframe}) an analysis which is very similar to
the previous one yields that

\begin{eqnarray}
-\,\frac{1}{\kappa}\,s^{*}\left(e^{\hat{r}}\,_{0}\,\sigma^{(+)}\,_{\hat{r}}\right)
& =  & \left\{ N\,\varepsilon |_{_{0}} - V^{b}\left(j_{b} |_{_{0}} -
\frac{i}{\kappa}\,\tau_{\hat{1}\hat{2}b}\right)\right\}\,\sqrt{\sigma}\,d^{2}x
\nonumber \\   & = & i\,\tilde{\xi}_{\hat{r}}\,^{a}\,{\cal
A}^{\hat{r}}\,_{a}\,d^{2}x\, , \label{SparlingAshHam} \end{eqnarray}
{\em where again the index} $\hat{r}$ {\em runs over the values}
$(\bot, \hat{a} = \hat{1},\hat{2})$.

\subsection{Off-shell expressions} Developments in the next section
devoted to the microcanonical action require an additional remark
concerning the Sparling forms.  The special gauge choices
(\ref{distinguishedframe}) and (\ref{distinguishedgauge}) may be
enforced on spacetime region $X$ {\em without} requiring that the
spacetime metric on $X$ is necessarily Ricci flat. In this more general
case the following relations hold:

\begin{equation} s^{*}\left(e^{\hat{r}}\,_{0}\,\sigma_{\hat{r}}\right)
= \left(N\, \omega^{\hat{a}}\,_{\vdash\hat{a}} -
V^{b}\,\beta^{\hat{a}}\,_{b}\,\omega^{\bot}\,_{\vdash\hat{a}}\right)\,\sqrt{\sigma}\,d^{2}x\,
, \label{SparlingGoldberg} \end{equation}

\begin{equation}
s^{*}\left(e^{\hat{r}}\,_{0}\,\sigma^{(+)}\,_{\hat{r}}\right) = \left\{
N\, \omega^{\hat{a}}\,_{\vdash\hat{a}} - V^{b}
\left(\beta^{\hat{a}}\,_{b}\,\omega^{\bot}\,_{\vdash\hat{a}} +
i\,\tau_{\hat{1}\hat{2}b}\right)\right\} \sqrt{\sigma} d^{2}x\, .
\label{complexSparlingGoldberg} \end{equation} These expressions are
the ``off-shell" versions of (\ref{SparlingHam}) and
(\ref{SparlingAshHam}).

\section{Microcanonical action}

The aim of this final section is to present two new expressions for the
microcanonical action in spacetime covariant form. As mentioned in the
last paragraph of the introductory section, the microcanonical action
is employed in a path integral representation of the density of quantum
states associated with the type of self-gravitating system under
analysis.\cite{micro} The task of building these expressions requires
the detailed analysis of two closely related tetrad action principles
for general relativity. Before displaying the new forms for the
microcanonical action, this section first discusses the construction
and (3+1)-decomposition of both these actions. Goldberg has given a
similar presentation before in a discussion of a covariant variational
principle for the conventional Ashtekar variables.\cite{Goldberg} Since
Goldberg's discussion adopts different conventions for the fundamental
definitions of (pseudo)Riemannian geometry than those given in this
paper, a partial account of his work is presented here in a notation
consistent with the previous sections. The first tetrad action, which
is referred to as the Goldberg action, is shown to be equivalent in a
definite sense to the action (\ref{BYaction1}) presented in the
preliminary section. The second tetrad action, referred to as the
complex Goldberg action, differs from the first by a complex boundary
term. Goldberg has demonstrated that the canonical analysis of this
action leads naturally to the $\Sigma$ Ashtekar variables. It is shown
that a careful treatment of the boundary terms associated with the
complex Goldberg action suggests the subtraction term (\ref{bdsubterm})
used for the introduction of the boundary Ashtekar variables.

Most of the assumptions about the spacetime region $M$ and notational
conventions presented in the preliminary section of this work still
hold. However, the special assumption (\ref{distinguishedgauge}) that
the leaves of the hypersurface foliation $\Sigma$ are orthogonal to
$^{3}\!B$ is momentarily relaxed in the interest of generality. In
addition, to begin with no special frame conditions are imposed on the
spacetime tetrad. When necessary later, the foliation assumption and
conditions on the tetrad are enforced. With regard to frame
requirements, a few comments are in order. In the following discussion
it is necessary to deal with spacetime tetrads which obey the time
gauge and which reduce to the distinguished frame
(\ref{distinguishedframe}) on $^{3}\! B$. The global existence of such
tetrads depends crucially on the topology of the generic slice
$\Sigma$. In general, there exist no global sections of the orthonormal
frame bundle which satisfy these two conditions simultaneously.
Therefore, the specification of such a special frame on $M$ may require
the use of multiple sections and transition functions. The subtleties
involved with such an analysis are not relevant for the present
discussion and are ignored here. Finally, in this last section the
spacetime metric $g$ on $M$ is {\em not} assumed to satisfy the vacuum
Einstein equations.

\subsection{Construction of the Goldberg actions} The machinery
developed in the preceding section on Sparling forms allows the
construction of the Goldberg first order tetrad action. A brief
manipulation yields that $e^{\hat{\alpha}} \wedge \tau_{\hat{\alpha}} =
0$, where $\tau_{\hat{\alpha}}$ is given by (\ref{tauSparling}). The
next necessary ingredient is the following well known identity obeyed
by the Einstein three-form:

\begin{equation} -e^{\hat{\alpha}} \wedge
G_{\hat{\alpha}}\,^{\hat{\beta}}\,e^{*}_{\hat{\beta}} = \Re e^{\ast}\,
, \end{equation} where $ e^{\ast} = e^{\bot} \wedge e^{\hat{1}} \wedge
e^{\hat{2}} \wedge e^{\hat{3}}$ is the volume form of $M$. With these
facts, the fundamental relation (\ref{Sparlingrelation}), and the first
Cartan structure equation, one verifies that

\begin{equation} \Re e^{\ast} = -e^{\hat{\alpha}} \wedge
d\sigma_{\hat{\alpha}} = \omega^{\hat{\alpha}}\,_{\hat{\beta}} \wedge
e^{\hat{\beta}} \wedge \sigma_{\hat{\alpha}} + d\left(e^{\hat{\alpha}}
\wedge \sigma_{\hat{\alpha}}\right)\, .  \end{equation} The above
expression leads one to examine the following frame-dependent first
order action:

\begin{equation} S\left[ e^{\hat{\alpha}}\right] =
\frac{1}{2\kappa}\int_{M}\omega^{\hat{\alpha}}\,_{\hat{\beta}} \wedge
e^{\hat{\beta}} \wedge \sigma_{\hat{\alpha}}\, .
\label{Goldbergaction1} \end{equation} This action is closely related
to a first order action originally used by Einstein \cite{York}, but it
differs from the Einstein action by a tetrad expression which cannot be
written in terms of the metric. The form of the Goldberg action
(\ref{Goldbergaction1}) is inherently frame-dependent.  However, though
the Goldberg Lagrangian does not transform as a gauge scalar with
respect to transformations of frame, it is manifestly invariant under
spacetime diffeomorphisms which leave the boundary fixed and preserve
the boundary conditions, since it is expressed solely in the language
of differential forms.\cite{DiffGeom}

The complex Goldberg tetrad action appropriate for the introduction of
the slice Ashtekar variables is closely related to the Goldberg action.
Remarkably, it turns out that

\begin{equation} \Re e^{\ast} =
\omega^{(+)\hat{\alpha}}\,_{\hat{\beta}} \wedge e^{\hat{\beta}} \wedge
\sigma^{(+)}\,_{\hat{\alpha}} + d\left(e^{\hat{\alpha}} \wedge
\sigma^{(+)}\,_{\hat{\alpha}}\right)\, , \end{equation} where
$\omega^{(+)\hat{\alpha}}\,_{\hat{\beta}}$ and the complex
superpotential $\sigma^{(+)}\,_{\hat{\alpha}}$ are defined respectively
by (\ref{sd/asdconnection}) and  (\ref{sd/asdsigSparling}). An
identical expression holds for the corresponding antiself-dual case.
This result suggests the alternative action

\begin{equation} S^{(+)}\left[ e^{\hat{\alpha}}\right]
 = \frac{1}{2\kappa}\int_{M}\omega^{(+)\hat{\alpha}}\,_{\hat{\beta}}
 \wedge e^{\hat{\beta}} \wedge \sigma^{(+)}\,_{\hat{\alpha}}\, ,
 \label{Goldbergaction2} \end{equation} which differs from the Goldberg
action (\ref{Goldbergaction1}) by the addition of an imaginary pure
divergence. Setting

\begin{equation} S^{(+)} = S + \Delta S^{(+)}\, , \end{equation} one
finds that the term appended to the Goldberg action
(\ref{Goldbergaction1}) is

\begin{equation} \Delta S^{(+)} =  \frac{1}{2\kappa}\int_{M}d\left\{
e^{\hat{\alpha}} \wedge \left( \sigma_{\hat{\alpha}} -
\sigma^{(+)}\,_{\hat{\alpha}}\right)\right\}\, .  \label{deltaS1}
\end{equation} With the enforcement of the time gauge and the
kinematical conditions (\ref{distinguishedframe}) and
(\ref{distinguishedgauge}) on the three-boundary, an appeal to the
generalized Stokes' theorem yields that

\begin{equation} \Delta S^{(+)} =
-\frac{i}{2\kappa}\int^{t''}_{t'}d^{3}x\,\epsilon^{\hat{r}\hat{s}\hat{t}}\,\omega_{\hat{s}\hat{t}
j}\tilde{E}_{\hat{r}}\,^{j} -
\frac{i}{2\kappa}\int_{^{3}\!B}d^{3}x\,\epsilon^{\hat{r}\hat{s}\hat{t}}\,\tau_{\hat{s}\hat{t}
j}\tilde{\xi}_{\hat{r}}\,^{j}\, . \label{deltaS2} \end{equation} In the
above formula $\tilde{E}_{\hat{r}}\,^{j} = det|E^{\hat{s}}\,_{k}|
E_{\hat{r}}\,^{j}$ and $\omega_{\hat{r}\hat{s}j}$ represent
respectively the densitized triad and connection coefficients induced
on either $t'$ or $t''$. Further properties of the slice triad and its
associated connection coefficients are found below. Either the integral
over $t'$ or the integral over $t''$ above may be viewed as the
generating functional often employed in the passage to the conventional
Ashtekar variables.\cite{tetradgrav} Later, the Goldberg action will be
identified with the action (\ref{BYaction1}) use by Brown and York.
Therefore, the preceding formula also exhibits the origin of the
subtraction term (\ref{bdsubterm}) introduced earlier in order to pass
to the boundary Ashtekar variables. This term arises because, subject
to the gauge conditions (\ref{distinguishedframe}) and
(\ref{distinguishedgauge}), the time gauge spacetime connection induces
the intrinsic connection (\ref{bdconnection}) on $^{3}\! B$. Also, in
order to obtain the  $^{3}\! B$ integral in (\ref{deltaS2}), it should
be noted that $\epsilon^{\hat{r}\hat{s}\hat{t}} = -
\epsilon^{\vdash\hat{r}\hat{s}\hat{t} }$.

\subsection{(3+1)-form of the Goldberg action} It is necessary to
perform a (3+1)-decomposition of the Goldberg action in order to
carefully examine its $^{3}\! B$ boundary terms. To facilitate this
decomposition by removing superfluous gauge variables from the start,
the time gauge condition is enforced and is retained for the remainder
of this paper.\cite{tetradgrav} Again, the time leg of the tetrad
$e_{\bot}$ coincides with the hypersurface normal of the foliation
$\Sigma$.  For the moment the kinematical conditions
(\ref{distinguishedframe}) and (\ref{distinguishedgauge}) on $^{3}\! B$
are not enforced.  It follows that both the tetrad and cotetrad can be
expressed in terms of the orthonormal triads and cotriads on the
$\Sigma$ slices by

\begin{eqnarray} e_{\bot} = \frac{1}{N}\left(\frac{\partial}{\partial
t} - V^{j}\frac{\partial}{\partial x^{j}}\right) &  & e_{\hat{r}} =
E_{\hat{r}}\,^{j}\frac{\partial}{\partial x^{j}} \nonumber \\ e^{\bot}
= Ndt\hspace{2.55cm} &  & e^{\hat{r}} = E^{\hat{r}}\,_{j}\left( dx^{j}
+ V^{k}dt\right)\, . \label{timetetrad} \end{eqnarray} The lapse
function $N$ and the shift vector field $\vec{V}$ are the kinematical
pieces of the spacetime metric $g$. In coordinates adapted to the
foliation $\Sigma$ the metric takes the following ADM form:

\begin{equation} g_{\mu\nu}\, dx^{\mu} \otimes dx^{\nu} = - Ndt \otimes
Ndt + h_{ij}\left( dx^{i} + V^{i}dt\right) \otimes \left( dx^{j} +
V^{j}dt\right)\, .  \end{equation} Orthonormality implies that
$\delta_{\hat{r}\hat{s}} =
h_{ij}\,E_{\hat{r}}\,^{i}\,E_{\hat{s}}\,^{j}$ where
$\delta_{\hat{r}\hat{s}} = diag (1,1,1) = \delta^{\hat{r}\hat{s}}$, the
slice metric $h_{ij}$ is related to the cotriad by $h_{ij} =
\delta_{\hat{r}\hat{s}}\,E^{\hat{r}}\,_{i}\,E^{\hat{s}}\,_{j}$, and
$\sqrt{h} = (E) = det|E^{\hat{r}}\,_{j}|$.

The (3+1)-decomposition of the Goldberg action is achieved by first
discerning the appropriate expressions for the connection coefficients
determined by (\ref{timetetrad}) and then expanding the action in terms
of these expressions. Appealing to (\ref{notorsion}), one calculates
the following batch of connection coefficients:

\begin{eqnarray}
  &  & \omega^{\bot}\,_{\hat{s}\bot} = E_{\hat{s}}\left[\log N\right]
  \nonumber \\ &  & \omega^{\bot}\,_{\hat{r}\hat{s}} =
  -\frac{1}{N}\left(
  h_{ij}\,E_{(\hat{r}}\,^{i}\,\dot{E}_{\hat{s})}\,^{j} +
  \overline{D}_{(\hat{s}}V_{\hat{r})}\right)  \label{sliceconnection}
  \\ &  & \omega^{\hat{t}}\,_{\hat{r}\hat{s}} =
  E_{\hat{s}}\,^{j}E^{\hat{t}}\,_{i}\left(
  D_{j}E_{\hat{r}}\,^{i}\right) \nonumber \\ &  &
  \omega_{\hat{r}\hat{s}\bot} = \frac{1}{N}\left( h_{ij}\,E_{[\hat
  {r}}\,^{i}\,\dot{E}_{\hat{s} ]}\,^{j} + \overline{D}_{[\hat
  {s}}V_{\hat{r} ]} -
  V^{\hat{t}}\omega_{\hat{r}\hat{s}\hat{t}}\right)\, . \nonumber
\end{eqnarray} Again, the dot denotes partial time differentiation.
Expanding the Goldberg action (\ref{Goldbergaction1}) in terms of these
coefficients, one arrives at

\begin{eqnarray}
  & S & = -\frac{1}{2\kappa}\int_{M}d^{4}x\,N(E)\left(
  \omega^{\bot}\,_{\hat{r}\hat{s}}\,\omega_{\bot}\,^{\hat{s}\hat{r}} -
  \omega_{\bot}\,^{\hat{r}}\,_{\hat{r}}\,\omega^{\bot\hat{s}}\,_{\hat{s}}
  \right. \nonumber \\ &   & \left. \hspace{2cm} - 2
  \omega_{\hat{r}}\,^{\bot}\,_{\bot}\,
  \omega^{\hat{r}\hat{s}}\,_{\hat{s}} +
  \omega^{\hat{t}}\,_{\hat{r}\hat{s}}\,\omega^{\hat{s}\hat{r}}\,_{\hat{t}}
  -

\omega_{\hat{t}}\,^{\hat{r}}\,_{\hat{r}}\,\omega^{\hat{t}\hat{s}}\,_{\hat{s}}\right)\,
  . \label{Goldbergaction3} \end{eqnarray}

In anticipation of working with the Ashtekar variables in the canonical
form of the theory, the densitized triad $\tilde{E}_{\hat{r}}\,^{j} =
(E)E_{\hat{r}}\,^{j}$ is chosen as the configuration space variable.
With this choice of configuration space variable the next step is to
calculate $\partial \tilde{E}_{\hat{r}}\,^{j}/ \partial t$. One
accomplishes this task by using the identity

\begin{equation} \delta E_{\hat{r}}\,^{j} = (E)^{-1}\left(
\delta^{\hat{s}}_{\hat{r}} \,\delta^{j}_{k} -
\frac{1}{2}E^{\hat{s}}\,_{k}E_{\hat{r}}\,^{j}\right)\delta\tilde{E}_{\hat{s}}\,^{k}
\label{yuck} \end{equation} in tandem with the explicit form of the
time gauge connection coefficients. A brief manipulation yields

\begin{eqnarray} \partial\tilde{E}_{\hat{r}}\,^{j} / \partial t & = &
N(E) E^{\hat{s}j}\left( \omega_{\bot\hat{s}\hat{r}} +
\omega_{\hat{s}\hat{r}\bot} -
\delta_{\hat{s}\hat{r}}\,\omega_{\bot}\,^{\hat{t}}\,_{\hat{t}}\right)
\nonumber  \\
  &   &  +\, (E) E^{\hat{s}j} \left(
  V^{\hat{t}}\,\omega_{\hat{s}\hat{r}\hat{t}} +
  \delta_{\hat{s}\hat{r}}\,\overline{D}_{\hat{t}}V^{\hat{t}} -
  \overline{D}_{\hat{r}}V_{\hat{s}}\right)\, .  \label{Edot}
\end{eqnarray} Armed with the above equation and the expression for the
slice Ricci scalar $R$ in terms of the slice connection coefficients
$\omega_{\hat{r}\hat{s}\hat{t}}$, one finds after some calculation that
the action (\ref{Goldbergaction3}) takes the following form:

\begin{eqnarray} S & = & \int_{M}d^{4}x\,\left( \frac{1}{\kappa}\,
K^{\hat{r}}\,_{j}\,\frac{\partial\tilde{E}_{\hat{r}}\,^{j}}{\partial t}
-N {\cal H} - V^{j} {\cal H}_{j}\right) \nonumber \\
  &   & - \int_{^{3}\! B}d^{3}x\,\frac{1}{\kappa}\,\,
  ^{_{2}}n_{i}\left\{
  N\tilde{E}_{\hat{r}}\,^{i}\,\omega^{\hat{r}\hat{s}}\,_{\hat{s}} +
  V^{j}\left(
  \delta^{i}_{j}\,K^{\hat{r}}\,_{k}\tilde{E}_{\hat{r}}\,^{k} - K
^{\hat{r}}\,_{j}\tilde{E}_{\hat{r}}\,^{i}\right)\right\}\, ,
\label{(3+1)action1} \end{eqnarray} where $^{_{2}}n$ is the unit normal
of $B$ in $\Sigma$. In the equation above

\begin{equation} K^{\hat{r}}\,_{j}= -\omega^{\bot\hat{r}}\,_{\hat{s}}\,
E^{\hat{s}}\,_{j}\, .\label{hybridexcurv} \end{equation} The time gauge
definition of the extrinsic curvature (\ref{excurvature}) establishes
that $K^{\hat{r}}\,_{j}$ is the hybrid extrinsic curvature
$K^{\hat{r}}\,_{\hat{s}}\, E^{\hat{s}}\,_{j}$. Inspection of the
formula for $\omega^{\bot}\,_{\hat{r}\hat{s}}$ (\ref{sliceconnection})
shows that the above hybrid extrinsic curvature is a complicated
expression built from $N, V^{j}, \tilde{E}_{\hat{r}}\,^{j}$, and
$\partial \tilde{E}_{\hat{r}}\,^{j}/ \partial t$.  Also, in the above
equation ${\cal H}$ and ${\cal H}_{j}$ are given by

\begin{equation} {\cal H} = \frac{1}{2\kappa(E)}\left(
K^{\hat{r}}\,_{i}\,K^{\hat{s}}\,_{j} -
K^{\hat{r}}\,_{j}\,K^{\hat{s}}\,_{i}\right)\tilde{E}_{\hat{r}}\,^{j}\tilde{E}_{\hat{s}}\,^{i}
-\frac{(E)}{2\kappa}\, R \label{scalar} \end{equation}

\begin{equation} {\cal H}_{j} = \frac{1}{\kappa}\, D_{k}\left(
K^{\hat{r}}\,_{j}\tilde{E}_{\hat{r}}\,^{k} -
\delta^{k}_{j}\,K^{\hat{r}}\,_{i}\tilde{E}_{\hat{r}}\,^{i}\right)\, .
\label{vector} \end{equation}

{}From the appearance of the $\Sigma$ connection coefficient in the
(3+1)-form of
the action (\ref{(3+1)action1}), it is evident that the choice of the
time gauge does not fully remove the frame-ambiguity associated with
the Goldberg action (\ref{Goldbergaction1}). However, the residual
frame-dependence has been swept into a three-boundary term. To remove
this left-over ambiguity, one specifies the tetrad on $^{3}\! B$.
Implementation of the kinematical conditions (\ref{distinguishedframe})
and (\ref{distinguishedgauge}) allows the (3+1)-form of the Goldberg
action to be expressed as

\begin{eqnarray} S & = & \int_{M}d^{4}x\,\left( \frac{1}{\kappa}\,
K^{\hat{r}}\,_{j}\,\frac{\partial\tilde{E}_{\hat{r}}\,^{j}}{\partial t}
-N {\cal H} - V^{j} {\cal H}_{j}\right) \nonumber \\
  &   & + \,\frac{1}{\kappa} \int_{^{3}\! B} dt\, e^{\hat{r}}\,_{0}\,
  \sigma_{\hat{r}}\,\, . \label{(3+1)action2} \end{eqnarray} The above
result is obtained by appealing to the ``off-shell" expression
(\ref{SparlingGoldberg}). One should note that this is {\em not} a
canonical action since $K^{\hat{r}}\,_{j}$ is just a short-hand
expression (\ref{hybridexcurv}). This is the action obtained by
directly ``translating" the canonical form of the action
(\ref{BYaction1}) presented by Brown and York \cite{BY} into the
language of densitized triads.  The details of this prescription are
given below in a similar ``translation" of the microcanonical action.

\subsection{(3+1)-form of the complex Goldberg action}

In order to obtain the appropriate (3+1)-form of the complex Goldberg
action, it suffices to carefully examine the (3+1)-decomposition of the
boundary term (\ref{deltaS1}) and discern its effect on the previous
result for the chosen (3+1)-form of the Goldberg action. Using the no
torsion condition (\ref{notorsion}), one quickly demonstrates that
$d\sigma^{(+)}\,_{\hat{\alpha}} = d\sigma_{\hat{\alpha}}$. This result
along with another appeal to the no torsion condition allows $\Delta
S^{(+)}$ to be written as follows:

\begin{equation} \Delta S^{(+)} =
\frac{1}{2\kappa}\int_{M}\omega^{\hat{\alpha}}\,_{\hat{\beta}} \wedge
e^{\hat{\beta}} \wedge \left( \sigma^{(+)}\,_{\hat{\alpha}} -
\sigma_{\hat{\alpha}}\right)\, .  \end{equation} Extensive manipulation
of the previous expression which involves the gymnastics of
$\epsilon^{\hat{r}\hat{s}\hat{t}}$ identities allows one to verify that

\begin{equation} \Delta S^{(+)} = \frac{i}{2\kappa}\int_{M}d^{4}x\,
N(E)\,\epsilon^{\bot\hat{r}\hat{s}\hat{t}}\,
\omega_{\hat{r}\hat{s}}\,^{\hat{p}}\left( \omega_{\bot\hat{p}\hat{t}} +
\omega_{\hat{p}\hat{t}\bot} - \delta_{\hat{t}\hat{p}}\,
\omega_{\bot}\,^{\hat{m}}\,_{\hat{m}}\right)\, .  \end{equation} After
a long calculation which employs the relation (\ref{Edot}) and the
algebraic Bianchi identity $R^{\hat{p}}\,_{[\hat{r}\hat{s}\hat{t}]} =
0$ expressed in terms of the slice connection coefficients, one finds
that the above equation can be rewritten as

\begin{eqnarray} \Delta S^{(+)}  & = &
\frac{i}{2\kappa}\int_{M}d^{4}x\,\epsilon^{\bot\hat{r}\hat{s}\hat{t}}\,\omega_{\hat{r}\hat{s}j}\,\frac{\partial\tilde{E}_{\hat{t}}\,^{j}}{\partial
t} \nonumber \\
  &   & + \frac{i}{2\kappa}\int_{^{3}\! B}d^{3}x\,\, ^{_{2}}n_{j}\left(
  \epsilon^{\bot\hat{r}\hat{s}\hat{t}}\,
  \tilde{E}_{\hat{r}}\,^{j}\,\omega_{\hat{s}\hat{t}k}\,V^{k} -

(E)\,\epsilon^{\bot\hat{r}\hat{s}\hat{t}}\,\omega_{\hat{r}\hat{s}\hat{t}}\,V^{j}\right)\,
  . \label{yuck2} \end{eqnarray} Appending the preceding term to the
(3+1)-form of the Goldberg action (\ref{(3+1)action1}), one discovers
that in the leading ``p\.{q}" term of the complex Goldberg action the
``momentum" is

\begin{equation} \frac{1}{\kappa}\, K_{_{NEW}}\,^{\hat{r}}\,_{j} =
\frac{1}{\kappa}\left( K^{\hat{r}}\,_{j} +
\frac{i}{2}\,\epsilon^{\bot\hat{r}\hat{s}\hat{t}}\,\omega_{\hat{s}\hat{t}j}\right)
\, . \label{hybridexcurv2} \end{equation} Defining
$\omega^{\hat{r}}\,_{j} =
\frac{1}{2}\,\epsilon^{\bot\hat{r}\hat{s}\hat{t}}\,\omega_{\hat{s}\hat{t}j}
=
-\frac{1}{2}\,\epsilon^{\hat{r}\hat{s}\hat{t}}\,\omega_{\hat{s}\hat{t}j}$,
one introduces the triad version of the {\em Sen connection} on
$\Sigma$ as

\begin{equation} A^{\hat{r}}\,_{j} := -iK_{_{NEW}}\,^{\hat{r}}\,_{j}
=  \omega^{\hat{r}}\,_{j} - i K^{\hat{r}}\,_{j}\, .
\label{Senconnection} \end{equation} The Sen connection is {\em not}
the Ashtekar connection which is a canonical variable. The Sen
connection is the pull back of a complexified $SO(3)$ connection on the
bundle of orthonormal frames over $\Sigma$, and its properties are well
known.\cite{Ashtekar1,Sen}

If one enforces the gauge choices (\ref{distinguishedframe}) and
(\ref{distinguishedgauge}), then the expression (\ref{yuck2}) becomes

\begin{equation} \Delta S^{(+)} =
\frac{i}{2\kappa}\int_{M}d^{4}x\,\epsilon^{\bot\hat{r}\hat{s}\hat{t}}\,\omega_{\hat{r}\hat{s}j}\,\frac{\partial\tilde{E}_{\hat{t}}\,^{j}}{\partial
t} - \frac{i}{\kappa}\int_{^{3}\! B}d^{3}x\,
\sqrt{\sigma}\,\tau_{\hat{1}\hat{2} b}\, V^{b}\, . \label{yuck3}
\end{equation} Appending the above term to the special (3+1)-form of
the Goldberg action (\ref{(3+1)action2}), one discovers that the
(3+1)-form of the complex Goldberg action is

\begin{eqnarray} S^{(+)} & = & \int_{M}d^{4}x\,\left(
\frac{1}{\kappa}\,
K_{_{NEW}}\,^{\hat{r}}\,_{j}\,\frac{\partial\tilde{E}_{\hat{r}}\,^{j}}{\partial
t} -N {\cal H} - V^{j} {\cal H}_{j}\right) \nonumber \\
  &   & + \,\frac{1}{\kappa} \int_{^{3}\! B} dt\, e^{\hat{r}}\,_{0}\,
  \sigma^{(+)}\,_{\hat{r}}\,\, . \label{(3+1)action3} \end{eqnarray}
The above result follows from the ``off-shell" relation
(\ref{complexSparlingGoldberg}).

\subsection{Tetrad expressions for the microcanonical action}

The necessary machinery is now in place to present the new expressions
for the Brown-York microcanonical action in spacetime covariant form.
The microcanonical action is given in canonical form by \cite{micro}

\begin{equation} S_{mic} =  \int_{M}d^{4}x\,\left(
p^{ij}\,\frac{\partial h_{ij}}{\partial t} -N {\cal H} - V^{j} {\cal
H}_{j}\right)\, . \label{microaction1A} \end{equation} In terms of the
conventional ADM variables the scalar and vector constraints take the
form

\begin{equation} {\cal H} = \frac{2\kappa}{\sqrt{h}} \left\{  p^{ij}\,
p_{ij} - \frac{1}{2}\, (p^{i}\,_{i})^{2}\right\} -
\frac{\sqrt{h}}{2\kappa} R       \label{ADMscalar} \end{equation}

\begin{equation} {\cal H}_{j} =  - 2 D_{i}\, p^{i}\,_{j}\, .
\label{ADMvector} \end{equation} As described in \cite{micro}, in the
variational principle associated with the above action
(\ref{microaction1A}) the quantities $\sqrt{\sigma}\,\varepsilon$,
$\sqrt{\sigma}\, j_{b}$, and $\sigma_{ab}$ are fixed on $^{3}\! B$. One
can translate the above formula for the microcanonical action directly
into the language of densitized triads. First, the ADM $p^{ij}$ is
demoted to an expression depending on $N, V^{i}, h_{ij}$, and
$\partial\, h_{ij}/\partial t$. This is achieved by using equation
(\ref{slicemomentum1}) and considering the extrinsic curvature to be
given by the standard formula

\begin{equation} K_{ij} = \frac{1}{2N}\left( - \frac{\partial
h_{ij}}{\partial t} + D_{j}\, V_{i} + D_{i}\, V_{j}\right)\, ,
\end{equation} which holds for coordinates adapted to the foliation.
Next, one employs the identity

\begin{equation} \frac{\partial
h_{ij}}{\partial\tilde{E}_{\hat{s}}\,^{k}} = (E)^{-1}\left(
E^{\hat{s}}\,_{k}h_{ij} - E^{\hat{s}}\,_{i}h_{kj} -
E^{\hat{s}}\,_{j}h_{ki}\right) \end{equation} and considers $h_{ij}$ to
be a secondary quantity derived from $\tilde{E}_{\hat{r}}\,^{j}$. This
prescription yields the following action:

\begin{equation} S'_{mic} =  \int_{M}d^{4}x\,\left( \frac{1}{\kappa}\,
K^{\hat{r}}\,_{j}\,\frac{\partial\tilde{E}_{\hat{r}}\,^{j}}{\partial t}
-N {\cal H} - V^{j} {\cal H}_{j}\right)\, , \label{microaction1B}
\end{equation} where now the constraints are given by (\ref{scalar})
and (\ref{vector}). In the above equation $K^{\hat{r}}\,_{j}$ depends
upon $N, V^{j}, \tilde{E}_{\hat{r}}\,^{j}$, and $\partial
\tilde{E}_{\hat{r}}\,^{j}/ \partial t$. The preceding action is {\em
not} in canonical form.

Subject to the distinguished gauge conditions
(\ref{distinguishedframe}) and (\ref{distinguishedgauge}), the chosen
(3+1)-form of the Goldberg action (\ref{(3+1)action2}) suggests that
the Brown-York microcanonical action may be expressed in the following
spacetime covariant form:

\begin{equation} S'_{mic} =
\frac{1}{2\kappa}\int_{M}\omega^{\hat{\alpha}}\,_{\hat{\beta}} \wedge
e^{\hat{\beta}} \wedge \sigma_{\hat{\alpha}} -
\frac{1}{\kappa}\int_{^{3}\! B}\,\! ^{\bot}e^{\hat{r}} \wedge
\sigma_{\hat{r}}\, , \label{microaction2} \end{equation} where $^{\bot}
e^{\hat{r}} := \left\langle e^{\hat{r}}\, ,\partial/\partial
t\right\rangle\,dt$ and $r$ is a $^{3}\! B$ index taking the values
$(\bot\, , \hat{a})$. Also subject to the same restrictions, the
considerations regarding the complex Goldberg action suggest the
following closely related expression for the microcanonical action in
spacetime covariant form:

\begin{equation} S''_{mic}
 = \frac{1}{2\kappa}\int_{M}\omega^{(+)\hat{\alpha}}\,_{\hat{\beta}}
 \wedge e^{\hat{\beta}} \wedge \sigma^{(+)}\,_{\hat{\alpha}} -
 \frac{1}{\kappa}\int_{^{3}\! B}\,\! ^{\bot}e^{\hat{r}} \wedge
 \sigma^{(+)}\,_{\hat{r}}\, . \label{microaction3} \end{equation} One
should note the equivalence of the actions (\ref{microaction1B}) and
(\ref{microaction2}).  However, the action (\ref{microaction3})
possesses a slight modification. Namely, after being cast into
(3+1)-form, the ``momentum" (\ref{hybridexcurv2}) in its leading
``p\.{q}" term has a complex piece. Since both of the covariant tetrad
versions of the microcanonical action are written solely in the
language of differential forms, they are manifestly invariant under
spacetime diffeomorphisms which leave the boundary fixed and preserve
the boundary conditions.\cite{DiffGeom} In addition, they are also
invariant under transformations of tetrad which preserve the time gauge
and the distinguished frame (\ref{distinguishedframe}) on $^{3}\! B$.
Such frame transformations are local $SO(3)$ rotations of the triad on
the $\Sigma$ slices which on the $B$ slices become just local $SO(2)$
rotations of the dyad.

\section{Acknowledgments} I thank J. D. Brown, L. Dolan, and J. W. York
for valuable insights and helpful discussions. I am particularly
indebted to J. D. Brown, J. Kusnet, and J. W. York for proofreading
this manuscript. Research support was received from the National
Science Foundation, grant number PHY-8908741.

{\tenrm \begin{tabular}{|l||c|c|c|c|}           \hline
    & spacetime & hypersurface      & three-boundary      &
    two-boundary \\ &    $M$    & $\Sigma$ embedded & $^{3}\! B$
    embedded & $B$ embedded \\ &             & in $M$       & in
    $M$       & in $\Sigma$ \\ \hline \hline unit        &   &   &
&        \\ normal      &   & $u$ & $n$ & $^{_{2}} n = n$ \\ \hline
       &    &   &   &      \\ metric & $g_{\mu\sigma}$ & $h_{ij}$ &
$\gamma_{ij}$ & $\sigma_{ab}$ \\ \hline orthonormal &   &   &   &    \\
frame       & $e_{\hat{\alpha}} =
e_{\hat{\alpha}}\,^{\mu}\,\partial_{\mu}$ & $E_{\hat{r}} =
E_{\hat{r}}\,^{j}\,\partial_{j}$ & $\xi_{\hat{r}} =
\xi_{\hat{r}}\,^{j}\,\partial_{j}$ & $\beta_{\hat{a}} =
\beta_{\hat{a}}\,^{b}\,\partial_{b}$ \\ \hline ADM         &   &   &
&      \\ momentum    &   & $p^{ij}$ & $\pi^{ij}$ &    \\ \hline
densitized  &   &   &   &       \\ triad       &   &   &
$\Pi^{\hat{r}}\,_{j}$ &     \\ momentum    &   &   &   &
\\ \hline coordinate  &   &   &   &      \\ covariant   &
$\nabla_{\mu}$ & $D_{j}$ & ${\cal D}_{j}$ & $^{2}\nabla_{b}$    \\
derivative  &   &   &   &       \\ \hline orthonormal &   &   &
&       \\ covariant   & ${\overline{\nabla}}_{\mu}$ &
${\overline{D}}_{j}$ & ${\overline{{\cal D}}}_{j}$ &
$^{2}{\overline{\nabla}}_{b}$ \\ derivative  &   &   &   &
\\ \hline intrinsic   &   &   &   &        \\ curvature   &
$\Re_{\mu\nu\sigma\lambda}$ & $R_{ijkl}$ & ${\cal R}_{ijkl}$ &
\\ \hline extrinsic   &   &   &   &         \\ curvature   &   &
$K_{ij}$ & $\Theta_{ij}$ & $k_{ab}$   \\ \hline \end{tabular}} \\[.5cm]
{\bf Conventions Table}

Some spaces above have been left blank because they are either not
applicable or not needed. The symbol $\Re$ is used for the Riemann
tensor on $M$ and is not a tensor density. The unit normal for $^{3}\!
B$ embedded in $M$ is also the unit normal for B embedded in $\Sigma$
by virtue of the condition that $\left( u \cdot n \right) = 0$ on
$^{3}\! B$.


\begin{thebibliography}{99}


\bibitem{BMYW} J. W. York, Phys. Rev. {\bf D33}, 2092 (1986); H. W.
Braden, B. F. Whiting, and J. W. York, Phys. Rev. {\bf D36}, 3614
(1987); B. F. Whiting and J. W. York, Phys. Rev. Lett.  {\bf 61}, 1336
(1988); J. W. York, Physica {\bf A158}, 425 (1989); J. D. Brown, G. L.
Comer, E. A. Martinez, J. Melmed, B. F. Whiting, and J. W. York, Class.
Quantum Grav. {\bf 7}, 1433 (1990); H. W. Braden, J. D. Brown, B. F.
Whiting, and J. W. York, Phys. Rev. {\bf D42}, 3376 (1990); J. W. York
in {\em Conceptual Problems of Quantum Gravity}, edited by A. Ashtekar
and J. Stachel (Birkh\"{a}user, Boston, 1991); J. D. Brown, E. A.
Martinez, and J. W. York, Phys. Rev. Lett. {\bf 66}, 2281 (1991).

\bibitem{BY} J. D. Brown and J. W. York, Phys. Rev. {\bf D47}, 1407
(1993).

\bibitem{micro} J. D. Brown and J. W. York, Phys. Rev. {\bf D47}, 1420
(1993).

\bibitem{York} J. W. York, Found. Phys. {\bf 16}, 249 (1986).

\bibitem{Ashtekar1} A. Ashtekar, {\em Lectures on Non-perturbative
Canonical Gravity} (World Scientific Publishing Co. Pte. Ltd.,
Singapore, 1991).

\bibitem{Brown} J. D. Brown, unpublished notes on triad gravity.

\bibitem{tetradgrav} M. Henneaux, J.E. Nelson, and C. Schomblond, Phys.
Rev. {\bf D39}, 434 (1989).

\bibitem{Sen} A. Sen, J. Math. Phys. {\bf 22}, 1781 (1981).

\bibitem{DiffGeom} M. G\"{o}ckeler and T. Sch\"{u}cker, {\em
Differential Geometry, Gauge Theories, and Gravity} (Cambridge
University Press, New York, 1987).

\bibitem{Sparling} M. Dubois-Violette and J. Madore, Commun. Math.
Phys. {\bf 108}, 213 (1987).

\bibitem{MTW} C. W. Misner, K. S. Thorne, and J. A. Wheeler, {\em
Gravitation} (Freeman, San Francisco, 1973).

\bibitem{Goldberg} J. N. Goldberg, Phys. Rev. {\bf D37}, 2116 (1988).

\end{thebibliography}
\end{document}